\documentclass[12pt,a4paper,reqno]{article}

\usepackage[utf8]{inputenc}
\usepackage[letterpaper]{geometry}
\usepackage{xcolor}

\usepackage{mathrsfs}
\usepackage{amsmath}
\usepackage{amssymb}
\usepackage{setspace}
\usepackage{esint}
\usepackage[round]{natbib}
\usepackage[bookmarks=false,
 breaklinks=false,pdfborder={0 0 1},backref=section,colorlinks=false]
 {hyperref}
\hypersetup{hypertex,colorlinks,urlcolor=darkblue,citecolor=darkblue,linkcolor=darkblue,pdftex,hypertexnames=false,colorlinks,urlcolor=darkblue,citecolor=darkblue,linkcolor=darkblue}

\makeatletter

\usepackage[table]{xcolor}
\newcommand{\sig}{\cellcolor{purple!25}}
\newcommand\crule[3][black]{\textcolor{#1}{\rule{#2}{#3}}}

\newcolumntype{L}[1]{>{\raggedright\arraybackslash}p{#1}}
\newcolumntype{C}[1]{>{\centering\arraybackslash}p{#1}}
\newcolumntype{R}[1]{>{\raggedleft\arraybackslash}p{#1}}

\usepackage{setspace}

\usepackage{mathrsfs}
\usepackage{amsfonts}
\usepackage{bm}
\usepackage{verbatim}
\usepackage{wasysym}
\usepackage{dsfont}

\usepackage{booktabs}
\usepackage{tabularx}
\usepackage{siunitx} 
\usepackage{mathrsfs}
\usepackage{threeparttable}
\usepackage{lipsum}
\usepackage{ltablex}
\usepackage{array}
\usepackage{makecell}
\usepackage{caption}
\usepackage{multirow}
\usepackage{float}
\usepackage{graphicx}
\usepackage{soul}
\usepackage{rotating}
\usepackage{algorithm}
\usepackage{subfig}
\usepackage[normalem]{ulem}

\usepackage[hang,flushmargin]{footmisc}
\normalbaselines 

\definecolor{sangre}{rgb}{0.6,0.18,0.19}
\definecolor{dullmagenta}{rgb}{0.6,0,0.6}
\definecolor{darkblue}{rgb}{0,0,0.7}
\definecolor{verdeprofundo}{rgb}{0.02,0.376,0.031}
\definecolor{olivegreen}{RGB}{13, 165, 100}
\definecolor{chocolate}{RGB}{33,33,33}
\definecolor{lightcyan}{rgb}{0.6,1,1}
\definecolor{purple}{RGB}{147, 81, 209}

\ifx\pdfoutput\undefined
\else
\fi

\providecommand{\U}[1]{\protect\rule{.1in}{.1in}}

\newcommand\T{\rule{0pt}{2.7ex}}
\newcommand\B{\rule[-1.2ex]{0pt}{0pt}}

\newtheorem{theorem}{Theorem}[section]

\newtheorem{assumption}{Assumption}

\newtheorem{lemma}[theorem]{Lemma}

\@addtoreset{lemma}{Appendix B}

\numberwithin{equation}{section}

\newcommand{\Keywords}[1]{\par\noindent {\small{\em Keywords\/}: #1}} 
\newcommand{\JELclass}[1]{\par\noindent {\small{\em JEL classification\/}: #1}} 

\defcitealias{DieboldYilmaz2014}{DY}

\title{Uncovering Sparse Financial Networks \\ with Information Criteria}
\author{
Fu Ouyang\thanks{School of Economics, The University of Queensland, 39 Blair Dr, St Lucia, QLD 4067, Australia. Email: \href{mailto:f.ouyang@uq.edu.au}{f.ouyang@uq.edu.au}.} 
\and 
Thomas T. Yang\thanks{Research School of Economics, The Australian National University, Canberra, ACT 0200, Australia. Email: \href{mailto:tao.yang@anu.edu.au}{tao.yang@anu.edu.au}.}
\and
Wenying Yao\thanks{Corresponding author: Melbourne Business School, The University of Melbourne, 200 Leicester St, Carlton, VIC 3053, Australia. Email: \href{mailto:w.yao@mbs.edu}{w.yao@mbs.edu}.}}

\date{\today}

\makeatother

\begin{document}
\maketitle
\begin{abstract}
\begin{singlespace}
Empirical measures of financial connectedness based on Forecast Error Variance Decompositions (FEVDs) often yield dense network structures that obscure true transmission channels and complicate the identification of systemic risk. This paper proposes a novel information-criterion-based approach to uncover sparse and economically meaningful financial networks. By reformulating FEVD-based connectedness as a regression problem, we develop a model selection framework that consistently recovers the active set of spillover channels. We extend this method to generalized FEVDs to accommodate correlated shocks and introduce a data-driven procedure for tuning the penalty parameter using pseudo-out-of-sample forecast performance. Monte Carlo simulations demonstrate the approach's effectiveness in finite samples, as well as its robustness to approximately sparse networks and heavy-tailed errors. Applications to global stock markets, S\&P 500 sectoral indices, and commodity futures highlight the prevalence of sparse networks in empirical settings.
\end{singlespace}

\vspace{0.3cm}

\JELclass{C12, C14, C21, C52}

\vspace{0.3cm}

\end{abstract}

\newpage
\section{Introduction}

Modern financial markets are becoming ever more complex. Analysts, policymakers, and investors constantly sift through an immense stream of information, attempting to understand how shocks in one corner of the system reverberate through others. The growing speed of information flows and the increasing integration of global markets have amplified both the frequency and complexity of such transmission mechanisms. For policymakers and market participants alike, the central challenge lies in identifying the channels through which disturbances spread, as these pathways are often opaque. To uncover these channels and quantify their importance, researchers have increasingly turned to network models, which offer a powerful framework for representing financial linkages and tracing the diffusion of shocks across the system.

A large and growing literature has demonstrated the usefulness of network models for studying interconnectedness and systemic risk. \citet{BillioEtAl2012} show how econometric measures of interconnectedness can serve as early-warning indicators of systemic stress. The theoretical work by \citet{AcemogluEtAl2015} highlight how the structure of financial networks fundamentally shapes the amplification and propagation of shocks. \citet{GreenwoodEtAl2015} document empirically how common exposures and balance-sheet linkages can generate fire-sale spillovers in the banking system. Complementary measures of systemic importance include CoVaR \citep{TobiasBrunnermeier2016}, SRISK \citep{EngleBrownlees2016}, and systemic expected shortfall \citep[SES,][]{AcharyaEtAl2016}, among others. 
More recent research has extended these ideas to high-dimensional and time-varying settings. \citet{BarunikKrehlik2018} decompose connectedness into frequency components, distinguishing short- and long-term spillovers. \citet{BarigozziEtAl2025} develop a framework of factor network autoregressions that combines dimension reduction with network analysis, offering a scalable approach to capturing high-dimensional financial interconnectedness. Collectively, these studies underscore the central role of network-based methods in empirical finance and highlight the challenge of distinguishing economically meaningful connections from statistical noise.

Within this broad agenda, the work of \citet[][hereafter \citetalias{DieboldYilmaz2014}]{DieboldYilmaz2009, DieboldYilmaz2012, DieboldYilmaz2014} has been especially influential. Their series of papers introduce a connectedness framework built on \textit{forecast error variance decompositions} (FEVDs) from vector autoregressions (VARs), providing a tractable and intuitive measure of directional spillovers across financial institutions and markets. This framework has since become a cornerstone of empirical research on systemic risk and financial contagion. Its extensions have been widely applied to study various asset markets under different network configurations \citep[see][among others]{DemirerEtAl2018, GreenwoodEtAl2019, BostanciYilmaz2020, AndoEtAl2022}. By quantifying how much of the forecast error variance of one variable is explained by shocks to other variables, the \citetalias{DieboldYilmaz2014} methodology translates the abstract concept of network connectedness into a concrete, widely applicable empirical measure.

Despite its widespread adoption, the \citetalias{DieboldYilmaz2014} framework faces an important limitation. Because it is based on FEVDs from a VAR, the resulting connectedness matrices are inherently dense; every variable’s forecast error variance is mechanically decomposed into contributions from shocks to all variables, so no entry is ever exactly zero in practice. This dense structure obscures the network's true architecture and complicates the identification of systemic vulnerabilities. This limitation is further aggravated by the fact that almost the entire \citetalias{DieboldYilmaz2014} literature has concentrated on point estimates of the connectedness measures, with their degree of statistical uncertainty unknown. This gap is not accidental. Deriving valid inference for FEVD-based connectedness measures is challenging because the decomposition depends nonlinearly on estimated VAR parameters and on the covariance structure of the shocks. Early work by \citet{Lutkepohl1990} derives the analytical expression of the FEVDs using the delta method. \citet{InoueKilian2002} later establish the validity of bootstrap methods for FEVDs, providing a more practical alternative to asymptotic approximations. Unfortunately, the asymptotically normal distribution of the FEVDs collapses at the boundary when the true connectedness is zero \citep{Lutkepohl1990}. As a result, studies that quantify statistical uncertainty within the \citetalias{DieboldYilmaz2014} framework have not yet appeared in the literature.

In practice, however, many of these estimated linkages are economically negligible, and interpreting them as meaningful channels of spillovers risks overstating the degree of interconnection in financial markets. Empirical evidence suggests that financial networks often exhibit sparse structures, in which a handful of connections dominate the transmission of shocks, whereas many others are weak or immaterial \citep[see, for example,][]{BillioEtAl2012, BarigozziBrownlees2019}. Capturing this sparsity is essential for both statistical efficiency and economic interpretation, as it highlights the critical pathways through which contagion and systemic risk emerge. 

To address this challenge, we propose an alternative, information-criterion-based approach that directly targets sparsity in FEVD-based network models and reconceptualizes it as a regression problem. In this view, the forecast error of a given variable is expressed as a linear combination of shocks, with coefficients (weights) that correspond directly to the FEVD elements. This reformulation allows us to draw on the rich literature on model selection using information criteria, such as AIC \citep{Akaike1974}, BIC \citep{Schwarz1978}, and their extensions. This framework penalizes unnecessary complexity and retains only the most relevant contributors to each forecast error variance, effectively shrinking negligible connectedness measures toward zero. Crucially, the information criterion approach provides a consistent model selection procedure: as the sample size grows, it converges to the true set of relevant spillover channels. In doing so, our method complements existing sparsity-inducing techniques while preserving the interpretability and tractability of the \citetalias{DieboldYilmaz2014} connectedness framework. At the same time, it offers a different perspective from popular shrinkage estimators such as LASSO-based VARs \citep{DemirerEtAl2018, GabauerEtAl2024} and Bayesian VARs with shrinkage priors \citep{KorobilisYilmaz2018}, both of which control overfitting by shrinking coefficients but do not guarantee consistent recovery of the true underlying network.

Building on this regression-based interpretation of FEVDs, our paper makes several contributions. First, we introduce an information criterion tailored to FEVD networks that consistently identifies economically meaningful spillover channels. Our approach systematically eliminates negligible links, thereby recovering the sparse topological structure of financial networks. Second, by framing the problem as a regression with orthogonal shocks, we establish a direct connection between the connectedness measures and modern model selection theory. This perspective not only clarifies the statistical underpinning of variance decomposition but also opens the door to well-developed tools for penalizing unnecessary complexity. Third, we extend the methodology to the generalized FEVD (GFEVD) setting \citep{PESARAN1998}, which is widely used in applied research due to its ordering-invariance. Last but not least, we provide a data-driven procedure for tuning the penalty parameter using pseudo-out-of-sample (POOS) forecasts, thereby balancing model fit and interpretability. These innovations together allow us to derive parsimonious and economically meaningful network structures.

We evaluate the finite-sample performance of the proposed information criteria through extensive Monte Carlo experiments across a range of data-generating processes (DGPs). The results demonstrate that the approach consistently and accurately identifies the true active and inactive spillover channels in both small and large networks. This consistency holds even under heavy-tailed error distributions and in approximately sparse settings where many connections are economically negligible. Our data-driven tuning procedure successfully balances model fit with parsimony, ensuring that the recovered network structure remains robust to sampling noise. The practical utility of our approaches is demonstrated through three empirical applications: (1) The network of global equity markets studied by \citet{DieboldYilmaz2009}; (2) The network of the S\&P500 sectoral indices; (3) The volatility network among a large panel of commodity futures. 
These applications confirm that uncovering sparsity provides a clearer, more interpretable map of financial interconnectedness, facilitating more effective systemic risk monitoring.

The remainder of the paper is organized as follows. Section \ref{sec:model} introduces the regression-based reformulation of FEVD and GFEVD, proposes the associated information criteria, establishes their consistency, and details the data-driven tuning parameter selection using POOS forecasts. Section \ref{sec:MC} reports the results of extensive Monte Carlo simulations. Section \ref{sec:application} presents three empirical applications to global stock markets, S\&P 500 sectors, and commodity futures. Section \ref{sec:conclusion} concludes the paper. Technical proofs and additional simulation results are collected in the Appendix.

\par\bigskip
\noindent\textbf{Notation.} All vectors are column vectors. Bold lowercase letters represent vectors ($\mathbf{a}$), bold uppercase letters represent matrices ($\mathbf{A}$), and non-bold lowercase letters represent scalars ($a$). Script uppercase letters represent sets ($\mathscr{M}$). For any two sets $\mathscr{M}$ and $\mathscr{N}$, $\mathscr{M}\cap \mathscr{N}$ denotes their intersection, $\mathscr{M}^c$ is the complement of $\mathscr{M}$, and $\vert\mathscr{M}\vert$ is the cardinality of $\mathscr{M}$. The notation $\mathbf{I}_{m}$ denotes the $m\times m$ identity matrix, $\textrm{diag}(a_1,\, \ldots, \,a_m)$ denotes the diagonal matrix with diagonal entries $a_1,\, \ldots, \,a_m$.  For a generic positive number $a$, $\lfloor a\rfloor$ and $\lceil a\rceil$ represent the largest integer less than or equal to $a$ and the smallest integer greater than or equal to $a$, respectively. We use $\Pr(\cdot)$ to denote probability. For any (stochastic) positive sequences $\{a_{n}\}$ and $\{b_{n}\}$, $a_{n}=O(b_{n})$ ($O_{P}(b_{n})$) indicates that the sequence $a_{n}/b_{n}$ is bounded (in probability). 

\section{The Regression Perspective}\label{sec:model} 

The $m\times1$ vector $\mathbf{y}_{t}$ is assumed to follow a covariance stationary VAR($p$) model: 
\begin{equation}
\mathbf{y}_{t} =\mathbf{c}+ \sum_{l=1}^{p}\mathbf{\Phi}_{l} \mathbf{y}_{t-l} +\boldsymbol{\varepsilon}_{t},
\label{eq:model}
\end{equation}
where $\mathbf{c}$ is the $m\times1$ vector of intercept, and $\mathbf{\Phi}_{l}$, $l=1, \,\ldots, \,p$,  are $m\times m$ coefficient matrices. The $m \times 1$ vector of error term $\boldsymbol{\varepsilon}_{t}$ is a white noise process with covariance matrix $\textrm{Var} (\boldsymbol{\varepsilon}_{t}) =\mathbf{\Sigma}$. Without loss of generality, we normalize the variance of each element in $\mathbf{y}_{t}$ to unity, such that $\textrm{Var}(y_{t,i})=1\text{,}$ $i=1, \ldots, m$.

We can transform the the VAR($p$) process in \eqref{eq:model} into its vector moving average (VMA) representation: 
\begin{equation}
\mathbf{y}_{t} =\boldsymbol{\mu} +\sum_{l=0}^{\infty} \mathbf{\Psi}_{l} \boldsymbol{\varepsilon}_{t-l} =\boldsymbol{\mu} +\sum_{l=0}^{\infty} \mathbf{\Psi}_{l} \mathbf{P} \boldsymbol{\xi}_{t-l},
\label{eq:yt}
\end{equation}
where $\boldsymbol{\mu} =(\mathbf{I}_{m} -\sum_{l=1}^{p} \mathbf{\Phi}_{l})^{-1} \mathbf{c}$ is an $m\times1$ vector, $\mathbf{P}$ is a non-singular matrix such that $\mathbf{\Sigma} =\mathbf{P} \mathbf{P}'$, $\boldsymbol{\xi}_{t-l}=\mathbf{P}^{-1} \boldsymbol{\varepsilon}_{t-l}$, and $\boldsymbol{\xi}_{t}$ is the $m\times1$ vector of orthogonal
structural shocks with $\textrm{Var}\left( \boldsymbol{\xi}_{t} \right)=\mathbf{I}_{m}$. The VMA coefficient matrix $\mathbf{\Psi}_{l}$ are calculated recursively: 
\begin{equation}
\mathbf{\Psi}_{l} =\mathbf{\Phi}_{1} \mathbf{\Psi}_{l-1} +\mathbf{\Phi}_{2} \mathbf{\Psi}_{l-2} +\cdots +\mathbf{\Phi}_{p} \mathbf{\Psi}_{l-p} ,\qquad l=1, \,2, \,\ldots,
\label{eq:Psi}
\end{equation}
with $\mathbf{\Psi}_{0}=\mathbf{I}_{m}$ and $\mathbf{\Psi}_{l}=\mathbf{0}$, for $l<0$. The reduced-form error $\boldsymbol{\varepsilon}_{t}$ is mapped into orthogonal structural shocks via $\boldsymbol{\varepsilon}_{t} =\mathbf{P} \boldsymbol{\xi}_{t}$. The macroeconometric literature has proposed many identification strategies of the structural shocks, which can be used to obtain a uniquely identified $\mathbf{P}$. The Cholesky decomposition \citep{Sims1980} is one of such examples. In this paper, the mapping $\mathbf{P}$ is assumed to be given. This gives users the freedom to choose identification strategies for structural shocks.

The VMA($\infty$) coefficients in \eqref{eq:yt}, $\mathbf{\Psi}_{l} \mathbf{P}$, $l=1, \,2, \,\ldots$, are also known as the \textit{impulse response functions} (IRFs). They represent the response of $\mathbf{y}_{t+l}$ to a one-time shock in $\boldsymbol{\xi}_t$. We denote the sample counterpart of $\mathbf{P}$ as $\hat{\mathbf{P}}$, which is obtained by imposing the same set of identifying restrictions on the sample residual covariance matrix $\hat{\mathbf{\Sigma}}$.

\subsection{A Regression Interpretation of FEVD}

Based on the VMA representation in \eqref{eq:yt}, the dynamics for $\mathbf{y}_{t+1}$ can be expressed as 
\begin{align}
\mathbf{y}_{t+1} & =\boldsymbol{\mu} +\mathbf{\Psi}_{0} \mathbf{P} \boldsymbol{\xi}_{t+1} +\sum_{l=1}^{\infty} \mathbf{\Psi}_{l} \,\boldsymbol{\varepsilon}_{t+1-l} \nonumber \\
& =\boldsymbol{\mu} +\mathbf{\Psi}_{0} \mathbf{P} \boldsymbol{\iota}_{1} \xi_{t+1,1} +\mathbf{\Psi}_{0} \mathbf{P} \boldsymbol{\iota}_{2} \xi_{t+1,2} +\cdots +\mathbf{\Psi}_{0} \mathbf{P} \boldsymbol{\iota}_{m} \xi_{t+1,m} +\mathbf{f}_{t},
\label{eq:decomp1}
\end{align}
where $\underset{m \times 1}{\boldsymbol{\iota}_{j}} \equiv (0, \,\ldots, \,0, \underset{j\textrm{-th}}{\underbrace{1}} ,0, \,\ldots, \,0)^{\prime}$ is a selection vector with 1 at the $j$-th position and 0 elsewhere, the component 
$\mathbf{f}_{t} \equiv \sum_{l=1}^{\infty} \mathbf{\Psi}_{l} \boldsymbol{\varepsilon}_{t+1-l}$ contains elements that are in the information set at time $t$, and we make use of the expression
\[
\boldsymbol{\xi}_{t+1}=\boldsymbol{\iota}_{1}\xi_{t+1,1}+\boldsymbol{\iota}_{2}\xi_{t+1,2}+\cdots+\boldsymbol{\iota}_{m}\xi_{t+1,m}.
\label{eq:xi_t}
\]

Equation \eqref{eq:decomp1} expresses $\mathbf{y}_{t+1}$ as a linear combination of the $m$ orthogonal structural shocks, $\boldsymbol{\iota}_{1} \xi_{t+1,1}, \boldsymbol{\iota}_{2} \xi_{t+1,2}, \ldots, \boldsymbol{\iota}_{m} \xi_{t+1,m}$, and a residual term $\mathbf{f}_{t}$, which is known at time $t$ and is uncorrelated with the structural shocks at time $t+1$. The $i$-th element in $\mathbf{y}_{t+1}$, denoted as $y_{t+1,i}$,
is 
\begin{align}
y_{t+1,i} & =\boldsymbol{\iota}_{i}' \left( \boldsymbol{\mu} +\mathbf{\Psi}_{0} \mathbf{P} \boldsymbol{\xi}_{t+1} +\mathbf{f}_{t} \right) \nonumber \\
& =\mu_{i} +\varphi_{i1} \xi_{t+1,1} +\varphi_{i2} \xi_{t+1,2} +\cdots +\varphi_{im} \xi_{t+1,m} +\boldsymbol{\iota}_{i}' \mathbf{f}_{t},
\label{eq:yt+11}
\end{align}
where $\varphi_{ij}\equiv\boldsymbol{\iota}_{i}'\mathbf{\Psi}_{0}\mathbf{\mathbf{P}}\boldsymbol{\iota}_{j}$,
$i,j=1,\,\ldots,\,m$. Equation \eqref{eq:yt+11} can be viewed as a single-equation linear regression of $y_{t+1,i}$ on a set of mutually uncorrelated structural shocks. It implies that
\begin{equation}
\textrm{Var}\left(y_{t+1,i}\right)=\varphi_{i1}^{2}+\cdots+\varphi_{im}^{2}+\textrm{Var}\left(\boldsymbol{\iota}_{i}'\mathbf{f}_{t}\right),\label{eq:phiij}
\end{equation}
since $\boldsymbol{\varepsilon}_{t}$ is white
noise and $\textrm{Var}\left(\boldsymbol{\xi}_{t}\right)=\mathbf{I}_{m}$.
In other words, $\varphi_{ij}^{2}$ represents the amount in the variance
of $y_{t+1,i}$ that is explained by the $j$-th structural shock $\xi_{t+1,j}$.

The one-step-ahead FEVD for
$y_{t,i}$, $i=1,\,\ldots,\,m$, is defined as 
\begin{equation}
\theta_{ij}^{1}=\dfrac{\left(\boldsymbol{\iota}_{i}'\mathbf{\Psi}_{0}\mathbf{\mathbf{P}}\boldsymbol{\iota}_{j}\right)^{2}}{\boldsymbol{\iota}_{i}'\mathbf{\Psi}_{0}\mathbf{\Sigma}\mathbf{\Psi}_{0}'\boldsymbol{\iota}_{i}}=\dfrac{\varphi_{ij}^{2}}{\boldsymbol{\iota}_{i}'\mathbf{\Psi}_{0}\mathbf{\Sigma}\mathbf{\Psi}_{0}'\boldsymbol{\iota}_{i}},\qquad j=1,\,\ldots,\,m.\label{eq:theta1j}
\end{equation}
Clearly, $\theta_{ij}^{1}$ is strictly non-negative for any $i,j=1,\,\ldots,\,m$.
Some standard calculations yield $\sum_{j=1}^{m}\varphi_{ij}^{2}=\boldsymbol{\iota}_{i}'\mathbf{\Psi}_{0}\mathbf{\Sigma}\mathbf{\Psi}_{0}'\boldsymbol{\iota}_{i}$, and thus $\sum_{j=1}^{m}\theta_{ij}^{1}=1$. So, each $\theta_{ij}^{1}$
measures the percentage contribution from the $j$-th structural shock
to the one-step-ahead forecast variance of the $i$-th variable $y_{t+1,i}$.

Equation \eqref{eq:theta1j} reveals that $\theta_{ij}^{1}=0$ if and only if its corresponding $\varphi_{ij}=0$. Also note that $\varphi_{ii} \neq 0$ as $\textrm{Var}\left(\varepsilon_{t,i} \right)>0$ for any $i=1, \ldots, m$. Therefore, uncovering sparsity in the network structure $\left(\theta_{i1}^{1},\theta_{i2}^{1},\,\ldots,\,\theta_{im}^{1}\right)$ is equivalent to identifying zero elements in the coefficient set $\left(\varphi_{i1},\varphi_{i2},\,\ldots,\,\varphi_{im}\right)$.
Given the relationship in \eqref{eq:yt+11} where $y_{t+1,i}$ is
a linear function of $\xi_{t+1,j}$, $j=1,\,\ldots,\,m$, we can identify
the zero elements in $\left(\varphi_{i1},\varphi_{i2},\,\ldots,\,\varphi_{im}\right)$
using information criteria to decide if certain structural shocks
do not contribute to explaining variations in $y_{t+1,i}$. This will
balance the goodness of fit and model parsimony, which we will discuss next in a more general setting.

\subsection{Sparsity in $H$-step-ahead FEVD}\label{subsec:FEVD} 

In the general case of $H$-step-ahead FEVD where
$H\geq1$, we examine the $i$-th element in $\mathbf{y}_{t+H}$,
denoted as $y_{t+H,i}$. By recursively applying the decomposition
in \eqref{eq:decomp1} and \eqref{eq:yt+11} $H$ times, we obtain:
\begin{align}
y_{t+H,i}=\mu_{i}+ & \sum_{h=0}^{H-1}\boldsymbol{\iota}_{i}'\mathbf{\Psi}_{h}\mathbf{P}\boldsymbol{\iota}_{1}\xi_{t+H-h,1}+\sum_{h=0}^{H-1}\boldsymbol{\iota}_{i}'\mathbf{\Psi}_{h}\mathbf{P}\boldsymbol{\iota}_{2}\xi_{t+H-h,2}\nonumber \\
 & \qquad\qquad+\cdots+\sum_{h=0}^{H-1}\boldsymbol{\iota}_{i}'\mathbf{\Psi}_{h}\mathbf{P}\boldsymbol{\iota}_{m}\xi_{t+H-h,m}+\boldsymbol{\iota}_{i}'\mathbf{f}_{t},\label{eq:yt+h1}
\end{align}
where we abuse the notation a bit by writing $\mathbf{f}_{t}=\sum_{l=H}^{\infty}\mathbf{\Psi}_{l}\boldsymbol{\mathbf{\mathbf{\varepsilon}}}_{t+H-l}$
and $i=1,\ldots,m$. Analogous to \eqref{eq:theta1j}, the $H$-step-ahead
FEVD, $\theta_{ij}^{H}$, is defined as 
\[
\theta_{ij}^{H}=\dfrac{\sum_{h=0}^{H-1}\left(\boldsymbol{\iota}_{i}'\mathbf{\Psi}_{h}\mathbf{P}\boldsymbol{\iota}_{j}\right)^{2}}{\sum_{h=0}^{H-1}\left(\boldsymbol{\iota}_{i}'\mathbf{\Psi}_{h}\mathbf{\Sigma}\mathbf{\Psi}_{h}'\boldsymbol{\iota}_{i}\right)^{2}}=\dfrac{\varphi_{ij}^{2}}{\sum_{l=1}^{m}\varphi_{il}^{2}},\qquad i,j=1,\,\ldots,\,m,
\]
where we continue to use $\varphi_{ij}^{2}$ as in \eqref{eq:phiij} to denote the contribution of the $j$-th structural shock, $\xi_{t+1,j},\,\ldots,\,\xi_{t+H,j}$, to the variance of $y_{t+H,i}$:
\[
\varphi_{ij}^{2}\equiv\sum_{h=0}^{H-1}\left(\boldsymbol{\iota}_{i}'\mathbf{\Psi}_{h}\mathbf{P}\boldsymbol{\iota}_{j}\right)^{2}.
\]
Combined with \eqref{eq:yt+h1}, the variance of $y_{t+H,i}$ can be expressed
as 
\[
\textrm{Var}\left(y_{t+H,i}\right)=\varphi_{i1}^{2}+\varphi_{i2}^{2}+\cdots+\varphi_{im}^{2}+\textrm{Var}\left(\boldsymbol{\iota}_{i}'\mathbf{f}_{t}\right).
\]
Sparsity in the $H$-step-ahead FEVD set $(\theta_{i1}^{H},\dots,\theta_{im}^{H})$ is therefore equivalent to identifying zero elements in $\left(\varphi_{i1},\,\varphi_{i2},\,\ldots,\,\varphi_{im}\right)$.
As shown in \eqref{eq:yt+h1}, we can view $y_{t+H,i}$ as a linear
function of structural shocks $\xi_{t+H-h,j}$ with $i,\,j=1,\,\ldots,\,m$ and $h=0,...,H-1$.
Combining the $m$ equations together for $\mathbf{y}_{t+H}$, zero
elements in the FEVD matrix, 
\[
\boldsymbol{\Theta}^H = \left(\begin{array}{cccc}
\theta_{11}^{H} & \theta_{12}^{H} & \cdots & \theta_{1m}^{H}\\
\theta_{21}^{H} & \theta_{22}^{H} & \cdots & \theta_{2m}^{H}\\
\vdots & \vdots & \ddots & \vdots\\
\theta_{m1}^{H} & \theta_{m2}^{H} & \cdots & \theta_{mm}^{H}
\end{array}\right),
\]
correspond directly to zero coefficients in regression \eqref{eq:yt+h1} for all $m$ elements in $\mathbf{y}_{t+H}$. Since the diagonal elements of $\boldsymbol{\Theta}^H$ are always non-zero, we only investigate its off-diagonal elements. Given this relationship, we can use model selection tools on the coefficients in the regression \eqref{eq:yt+h1} to identify the zero elements in $\left\{ \varphi_{ij}\right\} $, $i,\,j=1,\,\ldots,\,m$, $i\neq j$.

To formalize the concept of sparsity in the FEVD matrix $\boldsymbol{\Theta}^H$,
we distinguish between the parameters that are truly non-zero and
those that are zero. We define the \textit{active set} 
\[
\mathscr{M}=\left\{ (i,j):\varphi_{ij}\neq0, \, i\neq j\right\}, 
\]
which collects the indices of coefficients associated with shocks
that contribute non-zero variance to $\mathbf{y}_{t+H}$. Its size
is denoted by $k^{*}=\left|\mathscr{M}\right|$. The complement, 
\[
\mathscr{M}^{c}=\left\{ (i,j):\varphi_{ij}=0, \, i\neq j\right\} ,
\]
is the \textit{inactive set}, containing coefficients that should be excluded. A good model selection procedure should be able to consistently recover the active set as the sample size $T$ grows. 


\subsection{Information Criterion}\label{subsec:IC} 

Using the intuition of BIC \citep{Schwarz1978}, we define the information criterion for network selection as 
\begin{equation}
\textrm{IC}_{\textrm{FEVD}}^{H} \left( k,\lambda_{T} \right) =2T\log \left( m -\sum_{l=1}^{k} \left(\hat{\varphi}^{ \left( l \right)} \right)^{2} \right) +k\lambda_{T},
\label{eq:IC_general}
\end{equation}
where $\{\hat{\varphi}^{(1)}, \dots, \hat{\varphi}^{(m^{2}-m)}\}$ represents the set of off-diagonal coefficient estimates $\{\hat{\varphi}_{ij}\}_{i \neq j}$ sorted in descending order of absolute magnitude. Here,
$m$ represents the total variance of the $m$ elements in $\mathbf{y}_{t+H}$, $\sum_{l=1}^{k}\left(\hat{\varphi}^{\left(l\right)}\right)^{2}$ captures the variance explained by the $k$ most dominant cross-variable FEVD elements, and $\lambda_{T}$ is a positive tuning parameter that governs the trade-off between model fit and parsimony.\footnote{$\sum_{i=1}^m \hat{\varphi}_{ii}^2$ captures the variance explained by each variable's own structural shocks. This component is always included in the information criterion.} 

The number of non-zero elements in the FEVD matrix $\boldsymbol{\Theta}^H$ is selected according to 
\begin{equation}
\hat{k}=\arg\min_{k=1,\ldots,m^{2}-m}\textrm{IC}_{\textrm{FEVD}}^{H}\left(k,\lambda_{T}\right).
\label{eq:IC_k}
\end{equation}
Coefficients beyond the largest $\hat{k}$ FEVD elements are set to zero. We set the new estimates of $\varphi_{ij}$ after the selection as
\begin{equation}
\tilde{\varphi}_{ij}^{2}=\begin{cases}
\hat{\varphi}_{ij}^{2},\;\textrm{ for those corresponding }\hat{\varphi}^{\left(l\right)},\;l=1,\ldots,\hat{k}\\
0,\;\textrm{ otherwise}
\end{cases}.
\label{eq:final_phi2}
\end{equation}
In this way, sparsity in $\boldsymbol{\Theta}^H$ is achieved by zeroing out coefficients beyond the top $\hat{k}$ FEVD components. Denote the estimated active and inactive sets using the rule in \eqref{eq:final_phi2} as 
\begin{equation}
\mathscr{\widetilde{M}}=\left\{ (i,j):\tilde{\varphi}_{ij}\neq0,i\neq j\right\} \quad\text{and}\quad\widetilde{\mathscr{M}}^{c}=\left\{ (i,j):\tilde{\varphi}_{ij}=0,i\neq j\right\}.
\label{eq:est_active}
\end{equation}
The following assumptions are required to ensure the consistency of this selection procedure.

\begin{assumption} \label{A:model} 
The data $\mathbf{y}_{t}$ follows the stationary model specified in \eqref{eq:model}. The innovation process $\boldsymbol{\varepsilon}{}_{t}$ is serially uncorrelated with positive definite covariance matrix $\mathbf{\Sigma}=\textrm{Var}\left(\mathbf{\mathbf{\boldsymbol{\varepsilon}}}_{t}\right)$, and the fourth moment of $\boldsymbol{\varepsilon}_{t}$ exists. 
\end{assumption}

\begin{assumption} \label{A:m} The cross-sectional dimension $m$ remains fixed as $T\rightarrow\infty$. \end{assumption}

\begin{assumption} \label{A:A1} 
The coefficient estimator satisfies the convergence rate
\[
\hat{\varphi}_{ij}^{2}-\varphi_{ij}^{2}=O_{P}\left(T^{-1/2}\right), \textrm{ for all } i,\,j=1,\ldots, m.
\]
\end{assumption}

We provide an estimator of $\hat{\varphi}_{ij}^{2}$ in Procedure \ref{proc1} and show in Appendix \ref{APP:low_level_condition} that the proposed estimator satisfies this condition.

\begin{theorem}\label{TH:main} Suppose Assumptions \ref{A:model}--\ref{A:A1} hold. Let $\lambda_{T}\rightarrow\infty$ and $\lambda_{T}/T\rightarrow0$
as $T\rightarrow\infty$. Then, the selection procedure \eqref{eq:IC_k}--\eqref{eq:est_active}, based on the  information criterion proposed
in \eqref{eq:IC_general}, is consistent in the
sense that 
\[
\Pr\left(\hat{k}=k^{*}\right)\rightarrow1\quad\text{ and }\quad\Pr\left(\mathscr{\widetilde{M}}=\mathscr{M}\right)\rightarrow1.
\]
\end{theorem} 
The proof of Theorem \ref{TH:main} is provided in Appendix \ref{sec:appd1}.

Given a sample of size $T$, $\left\{ \mathbf{y}_{1},\,\mathbf{y}_{2},\,\ldots,\,\mathbf{y}_{T}\right\} $, the procedure to estimate the FEVD matrix while imposing sparsity using the proposed information criterion is outlined below in Procedure \ref{proc1}. The algorithm outlined in Procedure \ref{proc1} begins by estimating a standard VAR model and computing the corresponding FEVD measures. The procedure’s central innovation lies in treating network identification as a model selection problem. By ranking the estimated pairwise spillovers by magnitude, the algorithm iteratively evaluates the proposed information criterion to determine the optimal number of active connections. Minimizing this criterion establishes a data-driven threshold that separates economically significant transmission channels from statistical noise. 

\begin{algorithm}[ht]
\floatname{algorithm}{Procedure} \vspace{0.3cm}

\begin{enumerate}
\item Normalize $\left\{ \mathbf{y}_{t}\right\} _{t=1}^{T}$ so that the
variance of each element is unity. 
\item Regress $\mathbf{y}_{t}$ on $\mathbf{y}_{t-1},\,\mathbf{y}_{t-2},\,\ldots,\,\mathbf{y}_{t-p}$, $t=p+1,\,\ldots,T$ to obtain the estimated VAR($p$) coefficient matrices $\hat{\mathbf{c}}$ and $\mathbf{\hat{\Phi}}_{l}$, $l=1,\,\ldots,\,p$. 
\item Use these estimates to calculate the fitted values and residuals,
\[
\mathbf{\hat{y}}_{t}=\mathbf{\hat{c}} +\textstyle\sum_{l=1}^{p} \mathbf{\hat{\Phi}}_{l} \mathbf{y}_{t-l}, \qquad \boldsymbol{\hat{\varepsilon}}_{t} = \mathbf{y}_{t} - \hat{\mathbf{y}}_{t}.
\]
The sample residual covariance matrix is then $\hat{\boldsymbol{\Sigma}} = \frac{1}{T-p}\sum_{t=p+1}^{T} \boldsymbol{\hat{\varepsilon}}_{t} \boldsymbol{\hat{\varepsilon}}_{t}^{\prime}$. Imposing the given identification restrictions on $\hat{\boldsymbol{\Sigma}}$ leads to $\hat{\mathbf{P}}$.
\item Obtain the VMA($\infty$) coefficient matrices $\hat{\mathbf{\Psi}}_{l}$,
$l=1,\,2,\,\ldots$, recursively using \eqref{eq:Psi}. 
\item Choose a value of $H$, and calculate 
\[
\hat{\varphi}_{ij}^{2}=\textstyle\sum_{h=0}^{H-1}\left(\boldsymbol{\iota}_{i}'\mathbf{\hat{\Psi}}_{h} \hat{\mathbf{P}} \boldsymbol{\iota}_{j}\right)^{2},\qquad i,j=1,\ldots,m,
\]
and rank the $\hat{\varphi}_{ij}^{2}$'s for $i\neq j$ in descending order as $\left\{ \left( \hat{\varphi}^{(1)} \right)^{2}, \,\ldots, \,\left( \hat{\varphi}^{(m^{2}-m)} \right)^{2} \right\}$. 
\item Choose a value of $\lambda_{T}$, construct the information criterion $\textrm{IC}_{\textrm{FEVD}}^{H}\left(k,\lambda_{T}\right)$ in \eqref{eq:IC_general} for $k=1,\,2,\,\ldots,\,m^{2}-m$ and obtain $\hat{k}$ that minimizes $\textrm{IC}_{\textrm{FEVD}}^{H}$. 
\item The remaining $\hat{\varphi}_{ij}^{2}$'s that are smaller than $\left(\hat{\varphi}^{(\hat{k})}\right)^2$ are set to zero as in \eqref{eq:final_phi2} to obtain the network structure with sparsity. This gives the final set of FEVD estimates $\tilde{\varphi}_{ij}^2$, $i,j=1,\ldots, m$, $i\neq j$. The diagonal elements $\hat{\varphi}_{ii}^2>0$ are unchanged, $i=1, \ldots, m$.
\end{enumerate}
\caption{Estimate FEVD with sparsity}
\label{proc1} 
\end{algorithm}

The information criterion in \eqref{eq:IC_general} requires the user to specify a value of the tuning parameter $\lambda_{T}$. A convenient, theoretically grounded choice is $\lambda_{T}=\log T$, mirroring the construction of the BIC. If the cross-sectional dimension $m$ is explicitly taken into account, one may instead use $\lambda_{T}=(\log T)/m$, as suggested by the asymptotic approximation in \eqref{eq:step2approx}. In practice, data-driven rules for specifying $\lambda_{T}$ can be constructed to accommodate specific characteristics of the data. We discuss such a procedure in the following section.

\subsection{Data-Driven Selection of the Tuning Parameter $\lambda_{T}$}\label{subsec:lambda} 

A key input of the information criterion is the choice of the tuning parameter $\lambda_{T}$, which governs the trade-off between sparsity and model fit when determining the active set of FEVD elements. We propose a data-driven approach to choose $\lambda_T$ based on the POOS mean squared forecast error (MSFE), as outlined in Procedure \ref{proc2}.


\begin{algorithm}[p]
\floatname{algorithm}{Procedure} \vspace{0.3cm}
\begin{enumerate}
    \item Select a set of $Q$ candidate tuning parameters $(\lambda_{T}^{(1)},...,\lambda_{T}^{(Q)})$. For each $\lambda_{T}^{(q)}$, $q=1,\,\ldots,Q$, calculate the POOS-MSFE following the steps below. 
        \begin{enumerate}
            \item Start with the training set of size $S<T$, $\left\{ \mathbf{y}_{1},\,\mathbf{y}_{2},\,\ldots,\,\mathbf{y}_{S}\right\}$: 
                \begin{enumerate}
                    \item Implement Procedure \ref{proc1} to obtain the off-diagonal FEVD estimates $\tilde{\varphi}_{ij}^{2}$, $i,j=1,\ldots,m$, $i\neq j$, and the diagonal estimates $\hat{\varphi}^2_{ii}>0$, $i=1, \ldots, m$. 
                    \item With the values $\tilde{\varphi}_{ij}^{2}$ obtained above, define
                    \[
                    \hat{\mathbf{1}}_{ij} \equiv \mathbf{1}[\tilde{\varphi}_{ij}^{2}>0],\quad\text{ for }\;i,j=1,\,\ldots,\,m, \, i\neq j,
                    \]
                    which signals whether the lags of the $j$-th shock jointly have an impact on $y_{t,i}$. Set $\hat{\mathbf{1}}_{ii}=1$ for $i=1, \, 2, \, \ldots, \, m$, as they are all non-zero.
                    \item Predict $\mathbf{y}_{S+1}$ using the VAR estimates obtained in step (a),
                    \[
                    \mathbf{\hat{y}}_{S+1}=\mathbf{\hat{c}} +\textstyle\sum_{l=1}^{p} \mathbf{\hat{\Phi}}_{l} \mathbf{y}_{S+1-l}, \qquad \boldsymbol{\hat{\varepsilon}}_{S+1} = \mathbf{y}_{S+1} - \hat{\mathbf{y}}_{S+1}.
                    \]
                    Obtain the predicted structural shocks $\hat{\boldsymbol{\xi}}_{S+1} = \mathbf{\hat{P}}^{-1} \hat{\boldsymbol{\varepsilon}}_{S+1}$.
                    \item Taking into account on the sparsity structure in $\hat{\mathbf{1}}_{ij}$'s in step (b), compute the one-step-ahead POOS forecast of $y_{S+1,i}$ for each $i=1,...,m$ as 
                    \begin{align*}
                    \tilde{y}_{S+1,i}= & \hat{\mu}_{i}+\sum_{j=1,j\neq i}^m \hat{\mathbf{1}}_{ij}\cdot\left(\boldsymbol{\iota}_{i}'\mathbf{\hat{\Psi}}_{0}\mathbf{\hat{P}}\boldsymbol{\iota}_{j}\hat{\xi}_{S+1,j}\right)+\sum_{j=1}^m \hat{\mathbf{1}}_{ij}\cdot\left(\sum_{h=1}^{H-1}\boldsymbol{\iota}_{i}'\mathbf{\hat{\Psi}}_{h}\mathbf{\hat{P}}\boldsymbol{\iota}_{j}\hat{\xi}_{S+1-h,j}\right).
                    \end{align*}
                    \textcolor{red}{
                    }
                
                    \item Calculate the one-step-ahead POOS forecast error, $\tilde{u}_{S+1,i}\equiv y_{S+1,i}-\tilde{y}_{S+1,i}$
                    for each $i=1,...,m$. 
                \end{enumerate}
            \item With a rolling window of size $S$ as the training sample, i.e., $\left\{ \mathbf{y}_{2}, \,\mathbf{y}_{3}, \,\ldots, \,\mathbf{y}_{S+1} \right\}$, repeat steps (a)\text{--}(e) in 1.1, and obtain the one-step-ahead POOS forecast errors $\tilde{\mathbf{u}}_{S+2}$. Continue the rolling-window scheme until the end of the training set is the observation from period $T-1$. 
            \item Collect all POOS forecast errors $\tilde{\mathbf{u}}_{S+1}$, $\ldots$, $\tilde{\mathbf{u}}_{T}$ and compute the POOS-MSFE: 
            \[
            \widehat{\textrm{MSFE}}_{\textrm{POOS}}\left(\lambda_{T}^{(q)}\right)=\frac{1}{T-S}\sum_{j=1}^{T-S}\left(\frac{1}{m}\sum_{i=1}^{m} \tilde{u}_{S+j,i}^{2} \right).
            \]
        \end{enumerate}
    \item The optimal tuning parameter $\lambda_{T}^{*}$ is then selected as the minimizer of POOS-MSFE among the candidates, i.e., 
    \[
    \lambda_{T}^{*}=\arg\min_{\lambda_{T}\in\{\lambda_{T}^{(1)},\ldots,\lambda_{T}^{(Q)}\}}\widehat{\textrm{MSFE}}_{\textrm{POOS}}\left(\lambda_{T}\right).
    \]
\end{enumerate}
\caption{Choosing $\lambda_{T}$ based on POOS-MSFE}
\label{proc2} 
\end{algorithm}

Given a sample of size $T$, $\left\{ \mathbf{y}_{1}, \dots, \mathbf{y}_{T} \right\}$, we partition the data into a training set and a validation set. The training set consists of the first $S$ observations, $\left\{ \mathbf{y}_{1}, \dots, \mathbf{y}_{S} \right\}$, where $S < T$.\footnote{In practice, a common approach is to set $S = \lfloor \alpha T \rfloor$ with $\alpha \in (0, 1)$. The training set typically comprises at least 70\% of the total observations (i.e., $\alpha \geq 0.7$).} We apply Procedure \ref{proc1} to the training data to estimate the sparse FEVD matrix, generate a one-step-ahead POOS forecast (error) for $\mathbf{y}_{S+1}$.\footnote{Note that in Step 1.1(d), we exclude $\hat{\xi}_{S+1,i}$ from the construction of the POOS forecast for $y_{S+1,i}$, even though its associated impulse response is inherently non-zero. This structural shock is recovered directly from the realized observation $y_{S+1,i}$ in 1.1(c). Consequently, it represents contemporaneous information unavailable at the forecast origin and would introduce look-ahead bias if included.} This process is repeated using a rolling-window scheme with a fixed window size $S$. The observations in the validation set $\left\{ \mathbf{y}_{S+1}, \dots, \mathbf{y}_{T} \right\}$ are utilized to calculate $T-S$ POOS forecast errors, which are then aggregated to compute the POOS-MSFE. Since different values of $\lambda_T$ yield distinct sparsity structures in the FEVD matrix, they result in different POOS-MSFE values. In practice, the optimal tuning parameter, $\lambda_{T}^{*}$ is selected as the value that minimizes the POOS-MSFE from a user-specified candidate set.



\subsection{Extension to Generalized FEVD}\label{subsec:GFEVD}

The methodology developed for the FEVD with orthogonal structural shocks extends naturally to the generalized FEVD (GFEVD) of \citet{KoopPesaranPotter1996} and \citet{PESARAN1998}. Unlike the traditional FEVD which requires orthogonal structural shocks for variance decomposition, GFEVD allows correlated shocks and is invariant to variable ordering. However, GFEVD requires normally distributed errors for meaningful interpretation. Both FEVD and GFEVD are popular in empirical research, each with its own strengths and limitations. Our method applies to both decompositions, although the regression intuition is most transparent under the orthogonal FEVD case.

The $H$-step-ahead GFEVD $\vartheta_{ij}^{H}$ for $y_{t+H,i}$ is defined as 
\begin{equation}
\vartheta_{ij}^{H}=\frac{\sigma_{jj}^{-1} \sum_{h=0}^{H-1} \left( \boldsymbol{\iota}_{i}' \mathbf{\Psi}_{h} \mathbf{\Sigma} \boldsymbol{\iota}_{j} \right)^{2}}{\sum_{h=0}^{H-1}\boldsymbol{\iota}_{i}'\mathbf{\Psi}_{h}\mathbf{\Sigma}\mathbf{\Psi}_{h}'\boldsymbol{\iota}_{i}} = \frac{\psi_{ij}^{2}}{\sum_{h=0}^{H-1}\boldsymbol{\iota}_{i}'\mathbf{\Psi}_{h}\mathbf{\Sigma}\mathbf{\Psi}_{h}'\boldsymbol{\iota}_{i}}, \qquad i,\,j=1, \,\ldots, m,
\label{eq:GFEVD}
\end{equation}
where $\boldsymbol{\Sigma} =\big[ \sigma_{ij} \big]_{i,j=1}^{m}$ and $\psi_{ij}^{2}\equiv \sigma_{jj}^{-1} \sum_{h=0}^{H-1} \left( \boldsymbol{\iota}_{i}' \mathbf{\Psi}_{h} \mathbf{\Sigma} \boldsymbol{\iota}_{j} \right)^{2}$. Crucially, $\vartheta_{ij}^{H}= 0$ if and only if its numerator component $\psi_{ij}^{2}= 0$. 

Starting with the simplest case of $H=1$, \eqref{eq:GFEVD} becomes
\[
\vartheta_{ij}^{1}=\frac{\sigma_{jj}^{-1} \left( \boldsymbol{\iota}_{i}' \mathbf{\Psi}_{0} \mathbf{\Sigma} \boldsymbol{\iota}_{j} \right)^{2}}{\boldsymbol{\iota}_{i}' \mathbf{\Psi}_{0} \mathbf{\Sigma} \mathbf{\Psi}_{0}' \boldsymbol{\iota}_{i}} \; \text{ and } \; \psi_{ij} =\sigma_{jj}^{-1/2} \boldsymbol{\iota}_{i}' \mathbf{\Psi}_{0} \mathbf{\Sigma} \boldsymbol{\iota}_{j}, \qquad i, \,j=1, \ldots, m,
\]
The coefficient $\psi_{ij}$ corresponds to the generalized IRF (GIRF) of \citet[][Equation (10)]{PESARAN1998}. Analogous to \eqref{eq:decomp1}, we can write $\mathbf{y}_{t+1}$ as 
\begin{align}
    \mathbf{y}_{t+1} & =\boldsymbol{\mu} +\mathbf{\Psi}_{0} \sum_{j=1}^m \boldsymbol{\iota}_{j} \varepsilon_{t+1,j} +\mathbf{f}_{t} \nonumber \\
    & =\boldsymbol{\mu} +\mathbf{\Psi}_{0} \sum_{j=1}^m \left( \boldsymbol{\iota}_{j} \varepsilon_{t+1,j} + \mathbf{\Sigma}\boldsymbol{\iota}_{j} \frac{ \varepsilon_{t+1,j}}{\sigma_{jj}} -\mathbf{\Sigma}\boldsymbol{\iota}_{j} \frac{ \varepsilon_{t+1,j}}{\sigma_{jj}} \right) +\mathbf{f}_{t} \nonumber \\
    & =\boldsymbol{\mu} +\sum_{j=1}^m \mathbf{\Psi}_{0} \mathbf{\Sigma}\boldsymbol{\iota}_{j} \frac{ \varepsilon_{t+1,j}}{\sigma_{jj}} +\mathbf{\Psi}_{0} \sum_{j=1}^m \left( \mathbf{I}_m - \sigma_{jj}^{-1} \mathbf{\Sigma} \right) \boldsymbol{\iota}_{j} \varepsilon_{t+1,j} +\mathbf{f}_{t} \nonumber \\
    & =\boldsymbol{\mu} +\sum_{j=1}^m \underbrace{\left( \frac{\mathbf{\Psi}_{0} \mathbf{\Sigma}\boldsymbol{\iota}_{j}}{\sqrt{\sigma_{jj}}} \right)}_{\text{GIRF}} \cdot \left( \frac{ \varepsilon_{t+1,j}}{\sqrt{\sigma_{jj}}} \right) +\mathbf{\Psi}_{0} \underbrace{\sum_{j=1}^m \left( \mathbf{I}_m - \sigma_{jj}^{-1} \mathbf{\Sigma} \right) \boldsymbol{\iota}_{j} \varepsilon_{t+1,j}}_{\boldsymbol{\varepsilon}_{t+1}^{\perp}} +\mathbf{f}_{t},
    \label{eq:decomp2}
\end{align}
where $\sigma_{jj}^{-1/2} \mathbf{\Psi}_{0} \mathbf{\Sigma}\boldsymbol{\iota}_{j}$, $j=1, \ldots, m$, becomes regression coefficients, $\boldsymbol{\varepsilon}_{t+1}^{\perp}$ denotes the additional terms left from the GIRF decomposition of $\mathbf{y}_{t+1}$, and $\mathbf{f}_{t} =\sum_{l=1}^{\infty}\mathbf{\Psi}_{l}\boldsymbol{\varepsilon}_{t+1-l}$ as in the previous case for FEVD with orthogonal shocks. Note that in general $\varepsilon_{t+1,j}$, $j=1, \ldots, m$, and $\boldsymbol{\varepsilon}_{t+1}^{\perp}$ are correlated, because the GFEVD allows for correlated shocks. As a result, the variance of all components in \eqref{eq:decomp2} will not add up to be the same as the variance of $\mathbf{y}_{t+1}$. 

For the general case of $H$-step-ahead GFEVD in \eqref{eq:GFEVD}, we can decompose $y_{t+H,i}$ as 
\begin{align}
    y_{t+H,i} & =\mu_{i} +\sum_{h=0}^{H-1} \left( \frac{\boldsymbol{\iota}_{i}' \mathbf{\Psi}_{h} \mathbf{\Sigma} \boldsymbol{\iota}_{1}}{\sqrt{\sigma_{11}}} \right) \cdot \left(\frac{\varepsilon_{t+H-h,1}}{\sqrt{\sigma_{11}}} \right) +\sum_{h=0}^{H-1} \left( \frac{\boldsymbol{\iota}_{i}' \mathbf{\Psi}_{h} \mathbf{\Sigma} \boldsymbol{\iota}_{2}}{\sqrt{\sigma_{22}}} \right) \cdot \left( \frac{\varepsilon_{t+H-h,2}}{\sqrt{\sigma_{22}}} \right) \nonumber \\
    & \qquad  +\cdots +\sum_{h=0}^{H-1} \left( \frac{\boldsymbol{\iota}_{i}' \mathbf{\Psi}_{h} \mathbf{\Sigma} \boldsymbol{\iota}_{m} }{\sqrt{\sigma_{mm}}} \right) \cdot \left( \frac{\varepsilon_{t+H-h,m}}{\sqrt{\sigma_{mm}}} \right) +\sum_{h=0}^{H-1} \boldsymbol{\iota}_{i}' \mathbf{\Psi}_{h} \boldsymbol{\varepsilon}_{t+H-h}^{\perp} +\mathbf{\iota}_{i}' \mathbf{f}_{t},
    \label{eq:decomp3}
\end{align}
where the same notation follows. 


Equations \eqref{eq:decomp2} and \eqref{eq:decomp3} interpret the GFEVD \eqref{eq:GFEVD} from the regression perspective. The numerator in the GFEVD element, $\psi_{ij}^2$, does not have an interpretation as ``percentage of variation explained'' as it does in the orthogonal case, due to the correlated shocks in the GFEVD. However, the procedure of using the information criterion to detect zeros in $\psi_{ij}^2$'s remains valid, because the validity of our approach does not rely on such interpretation.

Similar to the FEVD case, the information criterion for GFEVD is defined as 
\begin{equation}
\textrm{IC}_{\textrm{GFEVD}}^{H}\left(k,\lambda_{T}\right)=2\log\left( m -\sum_{l=1}^{k} \left( \hat{\psi}^{\left(l\right)} \right)^{2} \right) +k \lambda_{T},
\label{eq:IC_GFEVD}
\end{equation}
where $\{\hat{\psi}^{(1)}, \dots, \hat{\psi}^{(m^{2}-m)}\}$ represents the set of off-diagonal coefficient estimates $\{\hat{\psi}_{ij}\}_{i \neq j}$ sorted in descending order of absolute magnitude. 

The choice of $k$, the data-driven penalty parameter $\lambda_{T}$, and the corresponding structure of $\left\{ \tilde{\psi}_{ij}^{2} \right\}$ follows the same approach as outlined in previous sections.  Specifically, one can apply Procedure \ref{proc1} to compute $\tilde{\psi}_{ij}$ for $i,j=1,\dots,m$ using \eqref{eq:IC_GFEVD}.  Subsequently, Procedure \ref{proc2} is applied to select $\lambda_T$ based on POOS-MSFE, with $\tilde{\psi}_{ij}$ replacing $\tilde{\varphi}_{ij}$ in Steps 1.1(a) and 1.1(b). In this context, the calculation of $\hat{\boldsymbol{\xi}}_{S+1}$ in Step 1.1(c) is bypassed while all other calculations remain unchanged, and the one-step-ahead POOS forecast of $y_{S+1,i}$ is calculated using \eqref{eq:decomp3} rather than \eqref{eq:yt+h1}. Since the validity of the procedure is essentially identical to the FEVD case, separate technical proofs are omitted.

\section{Monte Carlo Experiments}\label{sec:MC}

We evaluate the finite-sample performance of the methods proposed in Section \ref{sec:model} through Monte Carlo simulations. We examine DGPs that follow VAR($p$) models with $\textbf{c}=\textbf{0}$, and with multivariate normal and heavy-tailed Student-$t$ errors. Our simulation study investigates a range of parameter values: the lag length $p\in\{1,\, 4\}$, the number of variables $m\in\{10,\, 20\}$, the FEVD and GFEVD horizon $H\in\{1,\, 5,\, 10\}$, and the sample size $T\in\{500,\, 1000,\, 2000\}$.

Sparsity in the network is imposed by setting the coefficient matrices $\mathbf{\Phi}_{l}$, $l=1,\, \ldots, \,p$, and the error covariance matrix $\mathbf{\Sigma}$ to block diagonal matrices with the same block structure. Specifically, for any pair of nodes $(i,j)$ belonging to different blocks (i.e., unconnected nodes), the corresponding matrix elements are set to zero. The nonzero elements of $\mathbf{\Phi}_{l}$ are randomly drawn from a uniform distribution, $U(-1,1)$. To ensure stationarity, we verify that all roots of the characteristic polynomial lie within the unit circle. If this condition is not met, all elements of $\mathbf{\Phi}_{l}$ are shrunk by a factor of 0.9 until stationarity is achieved.

To generate $\mathbf{\Sigma}$, we first construct an initial matrix $\mathbf{\Sigma}^{(0)} =\mathbf{I}_{m} +\boldsymbol{\rho} \boldsymbol{\rho}' -\textrm{diag}(\rho_{1}^{2},\, \ldots, \,\rho_{m}^{2})$, where the vector $\boldsymbol{\rho} =(\rho_{1},\, \ldots, \,\rho_{m})'$ consists of elements $\rho_{i} \sim U(-1,1)$. Analogously to $\mathbf{\Phi}_{l}$, the block structure is enforced on $\mathbf{\Sigma}^{(0)}$ by setting $\sigma_{ij}^{(0) }=\sigma_{ji}^{(0)} =0$ for any unconnected pair $(i,j)$. If $\mathbf{\Sigma}^{(0)}$ is positive definite, we set $\mathbf{\Sigma}= \mathbf{\Sigma}^{(0)}$. Otherwise, we iteratively update the matrix via $\mathbf{\Sigma}^{(r+1)} =\omega_{r+1} \mathbf{\Sigma}^{(r)} +(1-\omega_{r+1}) \mathbf{I}_{m}$ with $\omega_{0}=1$ and $\omega_{r+1} =0.9 \, \omega_{r}$, for $r=0, 1, 2, \ldots$, until the smallest eigenvalue is positive. The final positive-definite matrix is then set as $\mathbf{\Sigma}$.


We recursively generate $1000+T$ values of $\mathbf{y}_{t}$ using the coefficient matrices $\mathbf{\Phi}_{l}$, $l=1,\, \ldots, \,p$ and $\mathbf{\Sigma}$. The first 1,000 observations are discarded to mitigate the influence of initial values. We report the average \textit{correct discovery rates} (CDRs) for connected nodes ($\textrm{CDR}_1$), isolated nodes ($\textrm{CDR}_0$), and the overall network ($\textrm{CDR}_a$), defined as
\[
\textrm{CDR}_1=\frac{\vert \mathscr{\widetilde{M}} \cap \mathscr{M} \vert}{\left| \mathscr{M} \right|},\;\; \textrm{CDR}_0= \frac{\vert \widetilde{\mathscr{M}}^{c} \cap \mathscr{M}^{c} \vert}{\left| \mathscr{M}^{c} \right|}, \; \textrm{ and } \; \textrm{CDR}_a =\frac{\vert \widetilde{\mathscr{M}} \cap \mathscr{M} \vert +\vert \widetilde{\mathscr{M}}^{c} \cap \mathscr{M}^{c} \vert}{\left| \mathscr{M} \right|+\left| \mathscr{M}^{c} \right|}.
\]
All results are based on 1,000 replications. Here, $\mathscr{M}$ and $\mathscr{M}^{c}$ denote the true active and inactive sets, respectively, and $\widetilde{\mathscr{M}}$ and $\widetilde{\mathscr{M}}^{c}$ represent their estimates as defined in \eqref{eq:est_active}. These sets are estimated using the information criteria $\textrm{IC}_{\textrm{FEVD}}^{H} (k,\lambda_{T}^{*})$ from \eqref{eq:IC_general} and $\textrm{IC}_{\textrm{GFEVD}}^{H} \left(k,\lambda_{T}^{*}\right)$ from \eqref{eq:IC_GFEVD}. For both criteria, the tuning parameter $\lambda_{T}^{*}$ is selected via Procedure \ref{proc2} and its variant for $\textrm{IC}_{\textrm{GFEVD}}^{H}$ detailed in Section \ref{subsec:GFEVD}, using an initial training set of size $S = 0.9 T$. The candidate sets for $\lambda_{T}^{*}$ depend on the criterion and the horizon $H$:
\begin{itemize}
    \item For $\textrm{IC}_{\textrm{FEVD}}^{H}$, we select $\lambda_{T}^{*}$ from $\{c\log T/m:c=0.1, \dots, 0.6\}$ for $H=1$ and from $\{c\log T/m:c=1, \dots, 6\}$ for $H \in \{5, 10\}$.
    \item For $\textrm{IC}_{\textrm{GFEVD}}^{H}$, we select $\lambda_{T}^{*}$ from $\{c\log T/m:c=0.2, \dots, 0.7\}$ for $H=1$ and from $\{c\log T/m:c=2, \dots, 7\}$ for $H \in \{5, 10\}$.\footnote{Exploratory simulations indicate that to achieve satisfactory finite-sample performance, $\textrm{IC}_{\textrm{GFEVD}}^{H}$ generally requires a slightly larger tuning parameter (penalty) than $\textrm{IC}_{\textrm{FEVD}}^{H}$. In our simulations, we restricted the candidate sets to reduce computational cost. Despite asymptotical equivalence, we recommend that practitioners consider a wider range of candidate values in empirical applications to potentially improve finite-sample performance.}
\end{itemize}

\subsection{Small Networks}\label{subsec:MCsmall}

We first examine small networks ($m=10$) with a block-diagonal structure, where nodes are connected if and only if they belong to the same group. We consider two settings by varying the number of groups and the number of isolated nodes ($m_0$) where each isolated node is treated as a singleton group:
\begin{itemize}
\item[(S1)] Two isolated nodes ($m_0=2$). The remaining 8 nodes are partitioned into one group of 4 and two groups of 2. This structure yields an active set of size $\left| \mathscr{M} \right|=16$ and an inactive set of size $\left| \mathscr{M}^c \right|=74$.
\item[(S2)] Four isolated nodes ($m_0=4$). The remaining 6 nodes form a single group. This structure yields an active set of size $\left| \mathscr{M} \right|=30$ and an inactive set of size $\left| \mathscr{M}^c \right|=60$.
\end{itemize}

Table \ref{tab:sim_small} summarizes the CDRs for small networks across various configurations of lag length $p$, sample size $T$, and forecast horizon $H$. Panel A presents the results for FEVD using Cholesky decomposition to obtain orthogonal shocks, and Panel B reports the results using GFEVD. As the sample size expands, both $\textrm{CDR}_1$ and $\textrm{CDR}_0$ (and consequently $\textrm{CDR}_a$) exhibit monotonic convergence toward one across all specifications. This improvement is particularly pronounced for short horizons $H=1$, where the FEVD estimates are more sensitive to sampling variability. When $T$ reaches 2,000, the overall accuracy ($\textrm{CDR}_a$) is consistently high ($\geq 95\%$) across most specifications with moderately large horizons ($H\geq 5$). These results confirm that our approach effectively distinguishes between structural spillovers and sampling noise as $T \to \infty$, consistent with Theorem \ref{TH:main}.

\begin{table}[p]
  \centering
  \begin{threeparttable}
  \caption{Correct discovery rates for small networks\label{tab:sim_small}}
  \setlength{\tabcolsep}{4pt} 
  \begin{tabular*}{\textwidth}{@{\extracolsep{\fill}}lc *{3}{S[table-format=1.3]} *{3}{S[table-format=1.3]} *{3}{S[table-format=1.3]}}
    \toprule \hline
    \multicolumn{2}{l}{\textbf{Panel A}} & \multicolumn{3}{c}{$H = 1$} & \multicolumn{3}{c}{$H = 5$} & \multicolumn{3}{c}{$H = 10$}\T \\ \cmidrule(lr){3-5} \cmidrule(lr){6-8} \cmidrule(lr){9-11}
    DGP & $T$ & {CDR$_1$} & {CDR$_0$} & {CDR$_a$} & {CDR$_1$} & {CDR$_0$} & {CDR$_a$} & {CDR$_1$} & {CDR$_0$} & {CDR$_a$} \\ \midrule
    
    \multicolumn{2}{c}{$p = 1$} \\ \cmidrule(lr){1-2}
    \multirow{3}{*}{S1} 
      & 500  & 0.624 & 0.925 & 0.898 & 0.897 & 0.957 & 0.946 & 0.929 & 0.931 & 0.931 \\
      & 1000 & 0.681 & 0.930 & 0.908 & 0.939 & 0.963 & 0.959 & 0.958 & 0.942 & 0.945 \\
      & 2000 & 0.736 & 0.936 & 0.918 & 0.965 & 0.967 & 0.967 & 0.976 & 0.947 & 0.952 \\ \addlinespace
    \multirow{3}{*}{S2} 
      & 500  & 0.483 & 0.921 & 0.848 & 0.851 & 0.952 & 0.918 & 0.899 & 0.935 & 0.923 \\
      & 1000 & 0.543 & 0.930 & 0.865 & 0.914 & 0.963 & 0.946 & 0.943 & 0.946 & 0.945 \\
      & 2000 & 0.601 & 0.938 & 0.882 & 0.950 & 0.972 & 0.965 & 0.971 & 0.959 & 0.963 \\ \midrule
    
    \multicolumn{2}{c}{$p = 4$} \\ \cmidrule(lr){1-2}
    \multirow{3}{*}{S1} 
      & 500  & 0.637 & 0.922 & 0.896 & 0.844 & 0.951 & 0.932 & 0.918 & 0.903 & 0.906 \\
      & 1000 & 0.697 & 0.929 & 0.909 & 0.929 & 0.959 & 0.953 & 0.968 & 0.915 & 0.924 \\
      & 2000 & 0.745 & 0.933 & 0.916 & 0.965 & 0.967 & 0.967 & 0.986 & 0.923 & 0.934 \\ \addlinespace
    \multirow{3}{*}{S2} 
      & 500  & 0.523 & 0.919 & 0.853 & 0.756 & 0.911 & 0.859 & 0.873 & 0.886 & 0.881 \\
      & 1000 & 0.585 & 0.926 & 0.870 & 0.872 & 0.942 & 0.919 & 0.941 & 0.919 & 0.926 \\
      & 2000 & 0.646 & 0.932 & 0.884 & 0.935 & 0.956 & 0.949 & 0.976 & 0.929 & 0.945 \\ \midrule[\heavyrulewidth]
    
    \multicolumn{2}{l}{\textbf{Panel B}} & \multicolumn{3}{c}{$H = 1$} & \multicolumn{3}{c}{$H = 5$} & \multicolumn{3}{c}{$H = 10$} \\ \cmidrule(lr){3-5} \cmidrule(lr){6-8} \cmidrule(lr){9-11}
    DGP & $T$ & {CDR$_1$} & {CDR$_0$} & {CDR$_a$} & {CDR$_1$} & {CDR$_0$} & {CDR$_a$} & {CDR$_1$} & {CDR$_0$} & {CDR$_a$} \\ \midrule
    
    \multicolumn{2}{c}{$p = 1$} \\ \cmidrule(lr){1-2}
    \multirow{3}{*}{S1} 
      & 500  & 0.727 & 0.755 & 0.750 & 0.946 & 0.963 & 0.960 & 0.961 & 0.954 & 0.955 \\
      & 1000 & 0.782 & 0.781 & 0.781 & 0.972 & 0.972 & 0.972 & 0.980 & 0.964 & 0.967 \\
      & 2000 & 0.825 & 0.799 & 0.803 & 0.985 & 0.977 & 0.979 & 0.990 & 0.971 & 0.974 \\ \addlinespace
    \multirow{3}{*}{S2} 
      & 500  & 0.681 & 0.731 & 0.714 & 0.930 & 0.952 & 0.944 & 0.957 & 0.945 & 0.949 \\
      & 1000 & 0.737 & 0.745 & 0.742 & 0.962 & 0.962 & 0.962 & 0.979 & 0.955 & 0.963 \\
      & 2000 & 0.788 & 0.761 & 0.770 & 0.982 & 0.968 & 0.973 & 0.990 & 0.962 & 0.971 \\ \midrule
    
    \multicolumn{2}{c}{$p = 4$} \\  \cmidrule(lr){1-2}
    \multirow{3}{*}{S1} 
      & 500  & 0.747 & 0.740 & 0.741 & 0.928 & 0.931 & 0.930 & 0.967 & 0.861 & 0.880 \\
      & 1000 & 0.790 & 0.774 & 0.777 & 0.969 & 0.943 & 0.947 & 0.986 & 0.880 & 0.899 \\
      & 2000 & 0.828 & 0.796 & 0.801 & 0.986 & 0.955 & 0.960 & 0.993 & 0.907 & 0.922 \\ \addlinespace
    \multirow{3}{*}{S2} 
      & 500  & 0.722 & 0.715 & 0.717 & 0.892 & 0.900 & 0.897 & 0.961 & 0.844 & 0.883 \\
      & 1000 & 0.766 & 0.753 & 0.757 & 0.952 & 0.934 & 0.940 & 0.985 & 0.882 & 0.916 \\
      & 2000 & 0.813 & 0.767 & 0.782 & 0.979 & 0.946 & 0.957 & 0.995 & 0.905 & 0.935 \\
    \hline \bottomrule
  \end{tabular*}%
  \begin{tablenotes}[flushleft]
    \footnotesize
    \item[]\textit{Notes}: This table reports the CDRs for 10-dimensional VAR($p$) models. S1 features a 10-node network with an active set size of $\left| \mathscr{M} \right|=16$, comprising one group of 4 nodes, two groups of 2 nodes, and 2 isolated nodes. S2 features a 10-node network with $\left| \mathscr{M} \right|=30$, comprising one group of 6 nodes and 4 isolated nodes.
  \end{tablenotes}
  \end{threeparttable}
\end{table}%

The forecast horizon $H$ affects the trade-off between sparsity and model fit. As $H$ increases, $\textrm{CDR}_1$ improves substantially, reflecting the fact that longer-horizon FEVDs and GFEVDs aggregate impulse responses over time and therefore amplify persistent transmission channels. However, $\textrm{CDR}_0$ declines marginally for larger $H$ at the same time, indicating that longer horizons tend to pick up weaker, indirect spillovers. Furthermore, increasing the lag length from $p=1$ to $p=4$ tends to reduce the CDRs. Richer model dynamics increase parameter dimensionality and introduce additional estimation uncertainty, making it harder to distinguish true spillovers from noise. This result suggests that larger sample sizes are necessary for accurate selection in higher-order VARs.

Finally, comparing Designs S1 and S2 shows that a higher proportion of connected nodes weakens the finite-sample performance, particularly for detecting true connections. Both FEVD and GFEVD exhibit superior $\textrm{CDR}_1$ for DGP S1 compared to DGP S2. This performance gap persists across sample sizes, indicating that identifying the topology of denser networks is inherently more challenging. On the other hand, the ability to identify zero elements is largely comparable across the two DGPs. Among the two methods, GFEVD often yields higher $\textrm{CDR}_1$ than FEVD for all sample sizes and horizons. Conversely, FEVD exhibits a stronger tendency toward parsimony, yielding higher $\textrm{CDR}_0$, particularly with shorter horizons ($H\leq 5$). 
As the sample size increases, both methods converge to comparable performance levels across all configurations. Given the similarity between the two types of decompositions, we only report simulation results using GFEVD in subsequent sections due to its popularity in empirical applications, and leave the results using FEVD with Cholesky decomposition to Appendix \ref{sec:tabs_and_figs}. 




\subsection{Large Networks}\label{subsec:MClarge}

The next set of simulations examines the scalability of the proposed selection approaches by applying them to larger networks with $m=20$ nodes. This yields 380 possible pairwise connections within the network. We simulate four specifications of network structure that vary in cluster sizes and the number of isolated nodes:
\begin{enumerate}
\item[(L1)] Four isolated nodes ($m_{0}=4$), with the remaining 16 nodes partitioned into one group of 8, one group of 4, and two groups of 2. This results in an active set of size $\left| \mathscr{M} \right|=72$ and an inactive set of size $\left| \mathscr{M}^c \right|=308$.
\item[(L2)] Six isolated nodes ($m_{0}=6$), with the remaining 14 nodes partitioned into one group of 10 and one group of 4. This results in an active set of size $\left| \mathscr{M} \right|=102$ and an inactive set of size $\left| \mathscr{M}^c \right|=278$.
\item[(L3)] Ten isolated nodes ($m_{0}=10$), with the other 10 nodes partitioned into five groups of 2. This results in an active set of size $\left| \mathscr{M} \right|=10$ and an inactive set of size $\left| \mathscr{M}^c \right|=370$.
\item[(L4)] Twenty isolated nodes ($m_{0}=20$), representing a fully disconnected network. In this design, all $\boldsymbol{\Phi}_{l}$ matrices are diagonal with elements randomly drawn from $U(-1,1)$, and $\boldsymbol{\Sigma}$ is a diagonal matrix with elements drawn from $U(0.25,1)$. This yields an active set of size $\left| \mathscr{M} \right|=0$ and an inactive set of size $\left| \mathscr{M}^c \right|=380$. 
\end{enumerate}

\begin{table}[p]
  \centering
  \begin{threeparttable}
  \caption{Correct discovery rates for large networks with GFEVD\label{tab:sim_large}}
  \setlength{\tabcolsep}{4pt} 
  \begin{tabular*}{\textwidth}{@{\extracolsep{\fill}} lc *{3}{S[table-format=1.3]} *{3}{S[table-format=1.3]} *{3}{S[table-format=1.3]}}
    \toprule \hline
    & & \multicolumn{3}{c}{$H = 1$} & \multicolumn{3}{c}{$H = 5$} & \multicolumn{3}{c}{$H = 10$}\T \\ \cmidrule(lr){3-5} \cmidrule(lr){6-8} \cmidrule(lr){9-11}
    DGP & $T$ & {CDR$_1$} & {CDR$_0$} & {CDR$_a$} & {CDR$_1$} & {CDR$_0$} & {CDR$_a$} & {CDR$_1$} & {CDR$_0$} & {CDR$_a$} \\ \midrule
    
    \multicolumn{2}{c}{$p = 1$} \\ \cmidrule(lr){1-2}
    \multirow{3}{*}{L1} 
      & 500   & 0.690 & 0.779 & 0.762 & 0.913 & 0.970 & 0.959 & 0.947 & 0.961 & 0.958 \\
      & 1000  & 0.742 & 0.805 & 0.793 & 0.955 & 0.979 & 0.974 & 0.974 & 0.971 & 0.972 \\
      & 2000  & 0.787 & 0.827 & 0.820 & 0.979 & 0.984 & 0.983 & 0.988 & 0.977 & 0.979 \\ \addlinespace
    \multirow{3}{*}{L2} 
      & 500   & 0.679 & 0.765 & 0.742 & 0.899 & 0.966 & 0.948 & 0.943 & 0.959 & 0.955 \\
      & 1000  & 0.735 & 0.789 & 0.774 & 0.949 & 0.977 & 0.969 & 0.974 & 0.967 & 0.969 \\
      & 2000  & 0.783 & 0.810 & 0.803 & 0.974 & 0.983 & 0.980 & 0.987 & 0.977 & 0.979 \\ \addlinespace
    \multirow{3}{*}{L3} 
      & 500   & 0.755 & 0.800 & 0.799 & 0.918 & 0.983 & 0.981 & 0.926 & 0.973 & 0.972 \\
      & 1000  & 0.798 & 0.825 & 0.824 & 0.953 & 0.986 & 0.985 & 0.957 & 0.978 & 0.978 \\
      & 2000  & 0.831 & 0.845 & 0.845 & 0.973 & 0.988 & 0.988 & 0.976 & 0.982 & 0.982 \\ \addlinespace
    \multirow{3}{*}{L4} 
      & 500   & \text{--} & 0.801 & 0.801 & \text{--} & 0.972 & 0.972 & \text{--} & 0.949 & 0.949 \\
      & 1000  & \text{--} & 0.828 & 0.828 & \text{--} & 0.979 & 0.979 & \text{--} & 0.958 & 0.958 \\
      & 2000  & \text{--} & 0.842 & 0.842 & \text{--} & 0.981 & 0.981 & \text{--} & 0.964 & 0.964 \\ \midrule
    
    \multicolumn{2}{c}{$p = 4$} \\ \cmidrule(lr){1-2}
    \multirow{3}{*}{L1} 
      & 500   & 0.725 & 0.750 & 0.745 & 0.842 & 0.955 & 0.934 & 0.929 & 0.908 & 0.912 \\
      & 1000  & 0.769 & 0.794 & 0.790 & 0.927 & 0.974 & 0.965 & 0.972 & 0.935 & 0.942 \\
      & 2000  & 0.811 & 0.823 & 0.821 & 0.969 & 0.982 & 0.979 & 0.989 & 0.955 & 0.962 \\ \addlinespace
    \multirow{3}{*}{L2} 
      & 500   & 0.720 & 0.740 & 0.734 & 0.819 & 0.943 & 0.910 & 0.919 & 0.889 & 0.897 \\
      & 1000  & 0.769 & 0.780 & 0.777 & 0.916 & 0.969 & 0.955 & 0.972 & 0.932 & 0.942 \\
      & 2000  & 0.809 & 0.811 & 0.810 & 0.964 & 0.975 & 0.972 & 0.990 & 0.948 & 0.959 \\ \addlinespace
    \multirow{3}{*}{L3} 
      & 500   & 0.771 & 0.761 & 0.761 & 0.961 & 0.981 & 0.980 & 0.977 & 0.948 & 0.948 \\
      & 1000  & 0.810 & 0.810 & 0.810 & 0.987 & 0.986 & 0.986 & 0.993 & 0.961 & 0.962 \\
      & 2000  & 0.842 & 0.841 & 0.841 & 0.995 & 0.988 & 0.988 & 0.998 & 0.966 & 0.967 \\ \addlinespace
    \multirow{3}{*}{L4} 
      & 500   & \text{--} & 0.763 & 0.763 & \text{--} & 0.986 & 0.986 & \text{--} & 0.952 & 0.952 \\
      & 1000  & \text{--} & 0.817 & 0.817 & \text{--} & 0.990 & 0.990 & \text{--} & 0.963 & 0.963 \\
      & 2000  & \text{--} & 0.837 & 0.837 & \text{--} & 0.991 & 0.991 & \text{--} & 0.964 & 0.964 \\
    \hline \bottomrule
  \end{tabular*}%
  \begin{tablenotes}[flushleft]
    \footnotesize
    \item[]\textit{Notes}: This table reports the CDRs for 20-dimensional VAR($p$) models. L1 features a network with an active set size of $\left| \mathscr{M} \right|=72$, comprising one group of 8 nodes, one group of 4, two groups of 2, and 4 isolated nodes. L2 features a network with $\left| \mathscr{M} \right|=102$, comprising one group of 10 nodes, one group of 4, and 6 isolated nodes. L3 features a network with an active set size of $\left| \mathscr{M} \right|=10$, comprising 5 groups of 2 nodes and 10 isolated nodes. L4 features a null network with $\left| \mathscr{M} \right|=0$, representing a fully disconnected structure where all nodes are isolated.
  \end{tablenotes}
  \end{threeparttable}
\end{table}%

Table \ref{tab:sim_large} summarizes the results for the four DGPs described above using GFEVD.\footnote{The simulation results using Cholesky decomposition to calculate the FEVD are shown in Table \ref{tab:sim_large_FEVD}.} It is evident that the proposed method scales well with network size. Despite the much higher dimension of the parameter, the overall discovery rates remain high, especially for designs with strong sparsity L3 and L4. In these cases, $\textrm{CDR}_0$ is very close to one when $H \ge 5$ even in relatively small samples. In other words, when the true network contains few or no linkages between nodes, the criterion almost always recovers the correct inactive set. This finding is especially important for empirical applications, where networks with many weak pairwise connections are often interpreted as evidence of pervasive connectedness, which might be deemed spurious under the information criterion.

\begin{figure}[p]
\centering
\caption{The distribution of the selected constant $c^\ast$\label{fig:gfevd_tuning}}
\subfloat[DGP L1 with GFEVD\B]{\includegraphics[width=1\textwidth]{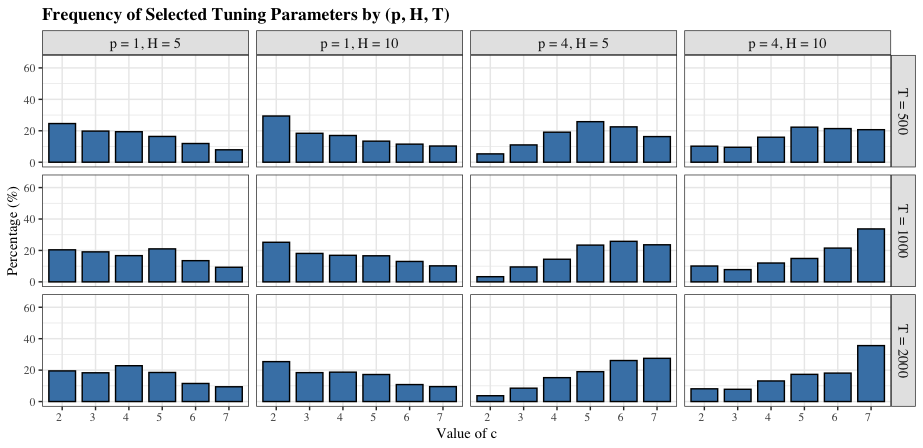}} \\ \vspace{0.5cm}
\subfloat[DGP L2 with GFEVD\B]{\includegraphics[width=1\textwidth]{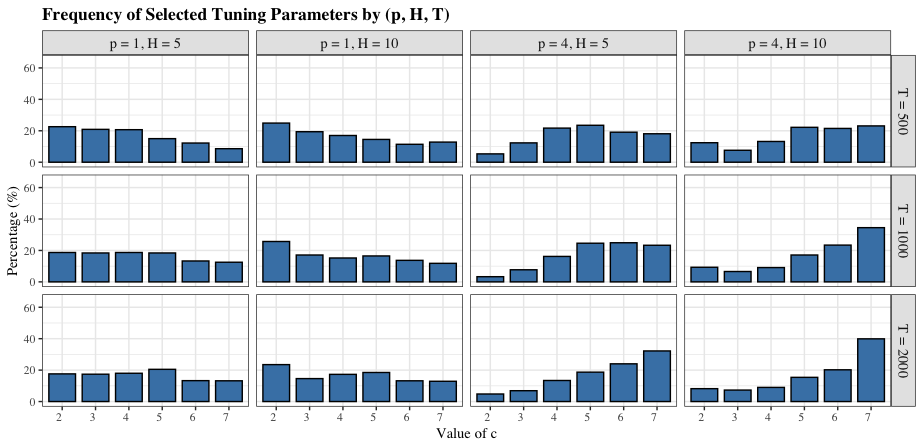}}
  \flushleft
    \footnotesize
    \item[]\textit{Notes}: These histograms present the frequencies of selecting $c$ in the tuning parameter $\lambda_T$ using Procedure \ref{proc2}. The candidate set for $c$ is from 1 to 7 with grid size 1.
\end{figure}

The overall results in Table \ref{tab:sim_large} exhibit similar patterns to those observed in Table \ref{tab:sim_small} for small networks. Higher values of $H$ lead to notable gains in $\textrm{CDR}_1$ but are accompanied by marginal declines in $\textrm{CDR}_0$. This trade-off is economically intuitive, as longer horizons reveal more connections at the cost of admitting some marginal linkages. It necessitates a larger penalty term in the information criteria for larger values of $H$ to maintain parsimony. This is indeed the case in the POOS-MSFE selection of the tuning parameter. To illustrate, Figure \ref{fig:gfevd_tuning} presents the distribution of the selected constant $c^\ast$ in the penalty term using Procedure \ref{proc2} for DGPs L1 and L2.\footnote{The distributions of $c^\ast$ for other settings are contained in Appendix \ref{sec:tabs_and_figs}.} These results reveal key patterns governing the optimal regularization strength. Across all designs and methods, increasing the forecast horizon $H$ or the lag order $p$ shifts the distribution of selected $c^\ast$ to higher values, suggesting larger penalties. This data-driven penalty selection approach ensures a reasonable balance between fit and parsimony. As a result, the overall $\textrm{CDR}_a$ in Table \ref{tab:sim_large} remains high across the board. These results confirm that the proposed approach is well-suited to high-dimensional settings and remains effective when the number of potential spillover channels grows.

\subsection{Approximately Sparse Networks}\label{subsec:MCdense}

Tables \ref{tab:sim_dense_fevd} and \ref{tab:sim_dense_gfevd} examine approximately sparse, or dense, networks in which all connections are non-zero but many are economically negligible. To simulate approximately sparse networks, we introduce two DGPs, D1 and D2, adapted from L1 and L2, respectively. Specifically, we replace all zeros in $\boldsymbol{\Phi}_{l}$ (i.e., those corresponding to unconnected node pairs) with small random numbers drawn from $U(-0.1,0.1)$. The error covariance matrix $\boldsymbol{\Sigma}$ is modified analogously while preserving its symmetry. We then apply the same iterative methods to adjust the $\boldsymbol{\Phi}_{l}$ and $\boldsymbol{\Sigma}$ matrices, ensuring the stationarity of the VAR($p$) process and the positive definiteness of $\boldsymbol{\Sigma}$. The resulting FEVD and GFEVD matrices contain no zero elements, implying that all nodes are technically connected. However, many elements remain very close to zero, representing rather weak connectivity. 

\begin{table}[ht]
  \centering
  \begin{threeparttable}
  \caption{Selection measures for 20-node approximately sparse networks with FEVD\label{tab:sim_dense_fevd}}
  \setlength{\tabcolsep}{4pt} 
  \begin{tabular*}{\textwidth}{@{\extracolsep{\fill}} lc *{3}{S[table-format=1.3]} *{3}{S[table-format=1.3]} *{3}{S[table-format=1.3]}}
    \toprule \hline
    &  & \multicolumn{3}{c}{$H = 1$} & \multicolumn{3}{c}{$H = 5$} & \multicolumn{3}{c}{$H = 10$} \\ \cmidrule(lr){3-5} \cmidrule(lr){6-8} \cmidrule(lr){9-11}
    DGP & $T$ & SP   & VL$_a$ & VL$_{o}$ & SP   & VL$_a$ & VL$_{o}$ & SP   & VL$_a$ & VL$_{o}$ \\ \midrule
    \multicolumn{2}{c}{$p = 1$} \\ \cmidrule(lr){1-2}
    \multirow{3}{*}{D1} 
          & 500   & 0.706 & 0.014 & 0.100 & 0.516 & 0.039 & 0.074 & 0.421 & 0.029 & 0.048 \\
          & 1000  & 0.654 & 0.007 & 0.055 & 0.406 & 0.022 & 0.041 & 0.319 & 0.015 & 0.025 \\
          & 2000  & 0.616 & 0.004 & 0.029 & 0.301 & 0.012 & 0.022 & 0.229 & 0.009 & 0.015 \\ \addlinespace
    \multirow{3}{*}{D2} 
          & 500   & 0.713 & 0.015 & 0.102 & 0.495 & 0.043 & 0.074 & 0.405 & 0.033 & 0.050 \\
          & 1000  & 0.666 & 0.008 & 0.057 & 0.390 & 0.023 & 0.039 & 0.304 & 0.016 & 0.024 \\
          & 2000  & 0.628 & 0.004 & 0.031 & 0.296 & 0.013 & 0.023 & 0.220 & 0.009 & 0.014 \\ \midrule
    \multicolumn{2}{c}{$p = 4$} \\  \cmidrule(lr){1-2}
    \multirow{3}{*}{D1} 
          & 500   & 0.696 & 0.012 & 0.088 & 0.666 & 0.067 & 0.185 & 0.559 & 0.055 & 0.124 \\
          & 1000  & 0.641 & 0.006 & 0.042 & 0.506 & 0.028 & 0.078 & 0.343 & 0.020 & 0.045 \\
          & 2000  & 0.602 & 0.003 & 0.020 & 0.361 & 0.013 & 0.035 & 0.199 & 0.008 & 0.018 \\ \addlinespace
    \multirow{3}{*}{D2} 
          & 500   & 0.712 & 0.013 & 0.089 & 0.612 & 0.074 & 0.180 & 0.503 & 0.055 & 0.111 \\
          & 1000  & 0.654 & 0.006 & 0.042 & 0.467 & 0.030 & 0.072 & 0.338 & 0.021 & 0.042 \\
          & 2000  & 0.612 & 0.003 & 0.019 & 0.342 & 0.013 & 0.032 & 0.207 & 0.009 & 0.018 \\
    \hline \bottomrule
  \end{tabular*}
  \begin{tablenotes}[flushleft]
    \footnotesize
    \item[]\textit{Notes}: This table presents the selection measures SP, VL$_a$ and VL$_o$ for 20-dimensional VAR($p$) models. DGPs D1 and D2 are adapted from L1 and L2, respectively, by replacing all zero elements in the coefficient and covariance matrices with small non-zero random numbers.
  \end{tablenotes}
  \end{threeparttable}
\end{table}

In this setting, traditional discovery metrics are no longer appropriate. Therefore, we report a set of sparsity and variance-loss measures. SP denotes the proportion of FEVD and GFEVD elements that are shrunk to zero using the information criterion. We also compute the total variation loss and the off-diagonal variation loss due to shrinkage. These measures are defined as follows: 
\[
    \text{SP} = \vert \tilde{ \mathscr{M} }^{c} \vert /(m^2-m), \quad
    \text{VL}_a = \dfrac{\sum_{(i,j) \in \tilde{\mathscr{M}}^{c}} \varphi_{ij}^{2} } {\sum_{i,j} \varphi_{ij}^{2}}, \quad \text{and} \quad
    \text{VL}_o = \dfrac{\sum_{(i,j) \in \tilde{ \mathscr{M} }^{c}} \varphi_{ij}^{2} } {\sum_{i\neq j} \varphi_{ij}^{2}}.
\]
Note that VL$_a$ and VL$_o$ are only interpretable for FEVD with orthogonal shocks. Therefore, we report the results using Cholesky factorization for the FEVD in Table \ref{tab:sim_dense_fevd} and relay the results using GFEVD with correlated shocks to Table \ref{tab:sim_dense_gfevd} in Appendix \ref{sec:tabs_and_figs}.

The results in Table \ref{tab:sim_dense_fevd} show that the information criterion can successfully enforce sparsity by shrinking a substantial fraction of weak linkages to zero, especially in smaller samples. More importantly, this sparsification incurs minimal cost in terms of explained variation. Both VL$_a$ and VL$_o$ are close to zero in all cases, and decline rapidly as the sample size $T$ increases, indicating that most of the eliminated connections contribute little to the total variance. Taken together, these results suggest that the proposed method serves as an effective filtering mechanism, removing negligible links while preserving the dominant transmission channels that drive the network dynamics.


\subsection{Heavy-tailed Errors}

\begin{table}[ht]
  \centering
  \begin{threeparttable}
  \caption{Correct discovery rates for 20-node networks with heavy-tailed errors and GFEVD}\label{tab:sim_heavy_tail}
  \setlength{\tabcolsep}{4pt}
  \begin{tabular*}{\textwidth}{@{\extracolsep{\fill}} lc *{3}{S[table-format=1.3]} *{3}{S[table-format=1.3]} *{3}{S[table-format=1.3]}}
    \toprule \hline
    & & \multicolumn{3}{c}{$H = 1$} & \multicolumn{3}{c}{$H = 5$} & \multicolumn{3}{c}{$H = 10$}\T \\ \cmidrule(lr){3-5} \cmidrule(lr){6-8} \cmidrule(lr){9-11}
    DGP & $T$ & {CDR$_1$} & {CDR$_0$} & {CDR$_a$} & {CDR$_1$} & {CDR$_0$} & {CDR$_a$} & {CDR$_1$} & {CDR$_0$} & {CDR$_a$} \\ \midrule
    
    \multicolumn{2}{c}{$p = 1$} \\ \cmidrule(lr){1-2}
    \multirow{3}{*}{H1} 
        & 500   & 0.719 & 0.548 & 0.580 & 0.867 & 0.910 & 0.902 & 0.917 & 0.885 & 0.891 \\
        & 1000  & 0.763 & 0.555 & 0.595 & 0.933 & 0.908 & 0.912 & 0.961 & 0.885 & 0.900 \\
        & 2000  & 0.807 & 0.553 & 0.601 & 0.966 & 0.915 & 0.925 & 0.981 & 0.896 & 0.912 \\ \addlinespace
    \multirow{3}{*}{H2} 
        & 500   & 0.711 & 0.535 & 0.582 & 0.854 & 0.894 & 0.883 & 0.917 & 0.870 & 0.883 \\
        & 1000  & 0.754 & 0.549 & 0.604 & 0.928 & 0.897 & 0.905 & 0.963 & 0.871 & 0.896 \\
        & 2000  & 0.801 & 0.545 & 0.613 & 0.967 & 0.899 & 0.917 & 0.984 & 0.880 & 0.908 \\ \midrule
    
    \multicolumn{2}{c}{$p = 4$} \\  \cmidrule(lr){1-2}
    \multirow{3}{*}{H1} 
        & 500   & 0.752 & 0.526 & 0.569 & 0.792 & 0.913 & 0.890 & 0.904 & 0.865 & 0.872 \\
        & 1000  & 0.791 & 0.538 & 0.586 & 0.908 & 0.921 & 0.919 & 0.966 & 0.887 & 0.902 \\
        & 2000  & 0.831 & 0.545 & 0.599 & 0.962 & 0.920 & 0.928 & 0.987 & 0.891 & 0.909 \\ \addlinespace
    \multirow{3}{*}{H2} 
        & 500   & 0.748 & 0.517 & 0.579 & 0.779 & 0.886 & 0.857 & 0.899 & 0.844 & 0.859 \\
        & 1000  & 0.791 & 0.528 & 0.598 & 0.898 & 0.905 & 0.903 & 0.967 & 0.864 & 0.892 \\
        & 2000  & 0.829 & 0.540 & 0.618 & 0.957 & 0.912 & 0.924 & 0.989 & 0.874 & 0.905 \\
    \hline \bottomrule
  \end{tabular*}%
  \begin{tablenotes}[flushleft]
    \footnotesize
    \item[]\textit{Notes}: This table presents the CDRs for 20-dimensional VAR($p$) models with heavy-tailed errors. DGPs H1 and H2 are adapted from L1 and L2, respectively, by replacing Gaussian errors with Student-$t$ errors.
  \end{tablenotes}
  \end{threeparttable}
\end{table}

To conclude our simulation study, we evaluate the performance of our approaches in VAR($p$) models with heavy-tailed errors. This is particularly relevant for financial network models, as a lot of financial market data are known to exhibit excess kurtosis and extreme observations. These characteristics may undermine methods that rely heavily on Gaussian assumptions of the innovations or are sensitive to outliers. We adapt DGPs L1 and L2 by replacing their Gaussian error distribution with a multivariate Student-$t$ distribution with $\nu=4$ degrees of freedom, while keeping the covariance matrix $\mathbf{\Sigma}$ and all other design aspects identical.\footnote{This is implemented using the \texttt{rmvt} function in the R package \texttt{mvtnorm}, with a scale matrix of $(\nu-2)/\nu\cdot \mathbf{\Sigma}$.} We label these designs as H1 and H2, respectively. Note that the analytical expression for GFEVD in Equation \eqref{eq:GFEVD} is derived using the conditional distribution of a multivariate Gaussian distribution. Therefore, it is technically incorrect when the innovations are from a Student-$t$ distribution.

The simulation results in Table \ref{tab:sim_heavy_tail} show that the ability to detect true connections remains consistently high under heavy-tailed errors.\footnote{The simulation results using Cholesky decomposition to calculate the FEVD are reported in Table \ref{tab:sim_heavy_tail_fevd}.} The CDR$_1$ for DGPs H1 and H2 is largely unaffected relative to the Gaussian benchmark in Table \ref{tab:sim_large} across all configurations. There are even slight improvements for short horizons $H=1$. This finding indicates that the proposed procedure reliably preserves economically meaningful spillover channels even when the data are contaminated by extreme shocks. Such properties are particularly desirable from an applied perspective, as missing key transmission links can lead to a serious underestimation of systemic risk or to the misidentification of influential nodes in the network.

At the same time, heavy-tailed errors lead to a modest reduction in CDR$_0$, especially when $H=1$, reflecting a higher tendency to retain some inactive links. Extreme observations inflate variance contributions and make weak or indirect spillovers more difficult to distinguish from genuinely relevant ones. As a result, the information criterion becomes slightly more conservative in pruning connections. From a practical standpoint, this trade-off might be desirable in financial applications. During turbulent periods, when extreme observations are more frequent, policymakers are often more concerned with avoiding false negatives than false positives. Retaining a small number of weak links is generally less costly than failing to identify key channels through which shocks propagate across the network. In this sense, the behavior of the proposed criterion under heavy-tailed errors aligns well with the priorities of systemic risk monitoring.



\section{Empirical Applications}\label{sec:application}

We use three examples to demonstrate the empirical relevance of the proposed information criteria for uncovering sparsity in financial network models. The first example replicates the global stock return network studied by \citet{DieboldYilmaz2009} based on FEVD. The second example investigates the sector indices of S\&P 500 constituent stocks using GFEVD. The last example analyzes the volatility network across a large panel of commodity futures using GFEVD.

\subsection{\citet{DieboldYilmaz2009} network}

Our first application uses the global stock market data from \citet{DieboldYilmaz2009} to demonstrate the use of the information criterion. This dataset comprises weekly returns on stock market indices across 19 markets from January 1992 to November 2007. A 19-variable VAR(2) is estimated on the full sample. The FEVD is based on $H=10$ and a Cholesky decomposition of shocks. The ordering of the variables is shown in Table \ref{tab:DY2009}, which reproduces Table 3 in \citet{DieboldYilmaz2009}. Each cell in Table \ref{tab:DY2009} shows the contribution of shocks to the country in the column heading to the 10-week-ahead forecast error variance of the country in the row heading. The off-diagonal elements represent cross-market linkages, while the diagonal elements represent own-market variation. Consistent with \citet{DieboldYilmaz2009}, we obtain a total spillover index of 35.5 percent, which represents the share of all cross-market contributions relative to the total forecast error variance across all markets. Two other summary metrics are also reported in Table \ref{tab:DY2009}: (1) The ``From spillover index'' (FIX) adds up each row excluding the diagonal element, measuring the total contribution of shocks received from all other nodes in the network; (2) The ``To spillover index'' (TIX) sums up each column excluding the diagonal element, measuring the each node's contribution to all other nodes' forecast error variance. 


\begin{sidewaystable}
\caption{\citet{DieboldYilmaz2009} global stock market return spillover table, 10/01/1992\text{--}21/11/2007 \label{tab:DY2009}}
\centering
\footnotesize 
\begin{tabular}{@{\hskip 0.5cm}l| rrrrrrrrrrrrrrrrrrr |r@{\hskip 0.5cm}}
\toprule \hline
& \multicolumn{19}{c|}{From} & \multirow{2}{*}{FIX}\T\B \\ \cline{2-20}
To  & US & UK & FR & DE & HK & JP & AU & ID & KR & MY & PH & SG & TW & TH & AR & BR & CL & MX & TR\T\B \\ \hline 
US  & 93.6\sig & 1.6\sig & 1.5\sig & 0.0  & 0.3  & 0.2  & 0.1  & 0.1  & 0.1  & 0.3  & 0.2  & 0.2  & 0.3  & 0.2  & 0.1  & 0.1  & 0.1  & 0.5  & 0.3 &  6\T \\ 
UK  & 40.3\sig & 55.8\sig & 0.7  & 0.4  & 0.1  & 0.5  & 0.1  & 0.2  & 0.1  & 0.3  & 0.2  & 0.0  & 0.1  & 0.1  & 0.1  & 0.1  & 0.0  & 0.4  & 0.6 & 44 \\ 
FR  & 38.3\sig & 21.7\sig & 37.2\sig & 0.1  & 0.0  & 0.2  & 0.3  & 0.3  & 0.3  & 0.2  & 0.2  & 0.1  & 0.1  & 0.2  & 0.1  & 0.1  & 0.1  & 0.1  & 0.3 &  63 \\ 
DE  & 40.8\sig & 15.9\sig & 13.0\sig & 27.6\sig & 0.1  & 0.1  & 0.3  & 0.4  & 0.6  & 0.1  & 0.3  & 0.3  & 0.1  & 0.2  & 0.0  & 0.1  & 0.0  & 0.1  & 0.1 &  72 \\ 
HK  & 15.3\sig & 8.7\sig & 1.7\sig & 1.4\sig & 69.9\sig & 0.3  & 0.0  & 0.1  & 0.0  & 0.3  & 0.1  & 0.0  & 0.2  & 0.9\sig & 0.3  & 0.1  & 0.1  & 0.3  & 0.4 & 30 \\ 
JP  & 12.1\sig & 3.0\sig & 1.8\sig & 0.9\sig & 2.3\sig & 77.7\sig & 0.2  & 0.3  & 0.3  & 0.1  & 0.2  & 0.3  & 0.3  & 0.1  & 0.1  & 0.0  & 0.0  & 0.1  & 0.1 & 22 \\ 
AU  & 23.2\sig & 6.0\sig & 1.3\sig & 0.2  & 6.3\sig & 2.3\sig & 56.9\sig & 0.1  & 0.4  & 0.1  & 0.2  & 0.2  & 0.5  & 0.5  & 0.1  & 0.3  & 0.1  & 0.6  & 0.7\sig &  43 \\ 
ID  & 6.0\sig & 1.6\sig & 1.2\sig & 0.7\sig & 6.4\sig & 1.6\sig & 0.4  & 77.0\sig & 0.7\sig & 0.5  & 0.1  & 0.9\sig & 0.2  & 1.0\sig & 0.7  & 0.1  & 0.3  & 0.1  & 0.3 & 23 \\ 
KR  & 8.3\sig & 2.6\sig & 1.3\sig & 0.7\sig & 5.7\sig & 3.7\sig & 1.0\sig & 1.2\sig & 72.8\sig & 0.0  & 0.1  & 0.1  & 0.1  & 1.3\sig & 0.2  & 0.2  & 0.1  & 0.1  & 0.7\sig &  27 \\ 
MY  & 4.0\sig & 2.2\sig & 0.6  & 1.3\sig & 10.5\sig & 1.5\sig & 0.4  & 6.6\sig & 0.5  & 69.2\sig & 0.1  & 0.1  & 0.2  & 1.1\sig & 0.1  & 0.6  & 0.4  & 0.2  & 0.3 &  31 \\ 
PH  & 11.1\sig & 1.6\sig & 0.3  & 0.2  & 8.1\sig & 0.4  & 0.9\sig & 7.2\sig & 0.1  & 2.9\sig & 62.9\sig & 0.3  & 0.4  & 1.5\sig & 1.6\sig & 0.1  & 0.1  & 0.1  & 0.2 &  37 \\ 
SG  & 16.8\sig & 4.8\sig & 0.7  & 0.9\sig & 18.5\sig & 1.3\sig & 0.4  & 3.2\sig & 1.6\sig & 3.6\sig & 1.7\sig & 43.1\sig & 0.3  & 1.1\sig & 0.8\sig & 0.5  & 0.1  & 0.3  & 0.4 &  57 \\ 
TW  & 6.4\sig & 1.3\sig & 1.2\sig & 1.8\sig & 5.3\sig & 2.8\sig & 0.4  & 0.4  & 2.0\sig & 1.0\sig & 1.0\sig & 0.9\sig & 73.6\sig & 0.4  & 0.8\sig & 0.3  & 0.1  & 0.3  & 0.0 &  26 \\ 
TH  & 6.3\sig & 2.5\sig & 1.0\sig & 0.7\sig & 7.8\sig & 0.2  & 0.8\sig & 7.6\sig & 4.6\sig & 4.0\sig & 2.3\sig & 2.2\sig & 0.3  & 58.2\sig & 0.5  & 0.2  & 0.1  & 0.4  & 0.3 &  42 \\ 
AR  & 11.9\sig & 2.1\sig & 1.6\sig & 0.1  & 1.3\sig & 0.8\sig & 1.3\sig & 0.4  & 0.4  & 0.6  & 0.4  & 0.6  & 1.1\sig & 0.2  & 75.3\sig & 0.1  & 0.1  & 1.4\sig & 0.3 & 25 \\ 
BR  & 14.1\sig & 1.3\sig & 1.0\sig & 0.7  & 1.3\sig & 1.4\sig & 1.6\sig & 0.6  & 0.5  & 0.7\sig & 1.0\sig & 0.8\sig & 0.1  & 0.7\sig & 7.1\sig & 65.8\sig & 0.1  & 0.6  & 0.7 &  34 \\ 
CL  & 11.8\sig & 1.1\sig & 1.0\sig & 0.0  & 3.2\sig & 0.6  & 1.4\sig & 2.3\sig & 0.3  & 0.3  & 0.1  & 0.9\sig & 0.3  & 0.8\sig & 2.9\sig & 4.0\sig & 65.8\sig & 2.7\sig & 0.4 &  34 \\ 
MX  & 22.2\sig & 3.5\sig & 1.2\sig & 0.4  & 3.0\sig & 0.3  & 1.2\sig & 0.2  & 0.3  & 0.9\sig & 1.0\sig & 0.1  & 0.3  & 0.5  & 5.4\sig & 1.6\sig & 0.3  & 56.9\sig & 0.6 &  43 \\ 
TR  & 3.0\sig & 2.5\sig & 0.2  & 0.7\sig & 0.6  & 0.9\sig & 0.6  & 0.1  & 0.6  & 0.3  & 0.6  & 0.1  & 0.9\sig & 0.8\sig & 0.5  & 1.1\sig & 0.6  & 0.1  & 85.8\sig & 14\B \\ \hline
TIX & 292 & 84 &  31 &  11 & 81 & 19 &  12 & 31 &  14 &  16 &  10 &   8 &   6 &  12 & 21 &   9 &   3 &   8 &  7 & 35.5\T\B \\ \hline
IN   & 2  & 1  & 2  & 3 & 5  & 5 & 6 & 9 & 10 & 7 & 8 & 11 & 11 & 11 & 8 & 11 & 11 & 9 & 7\T \\
OUT  & 18 & 18 & 13 & 9 & 13 & 9 & 7 & 6 & 4  & 6 & 5 & 5  & 2  & 9  & 6 & 3  & 0  & 2 & 2\B \\
\hline \bottomrule 
\end{tabular}
\vspace{0.2cm}

\footnotesize
\parbox{22.5cm}{\textit{Notes}: Numbers in this table replicate Table 3 of \citet{DieboldYilmaz2009}. The FEVD uses a VAR(2) for weekly stock market returns with Cholesky factorization and forecast horizon $H=10$, sample period 10/01/1992\text{--}21/11/2007. Cells shaded in \crule[purple!25]{0.6cm}{0.25cm} denote spillovers selected by the information criterion. The penalty term is determined using the POOF-MSFE procedure outlined in Section \ref{subsec:lambda}.}
\end{sidewaystable}

We use the methodology introduced in Section \ref{sec:model} and report the selected non-zero pairwise spillovers in Table \ref{tab:DY2009} in purple shades. The penalty term in the information criterion is determined using the POOF-MSFE procedure outlined in Section \ref{subsec:lambda}. Examining the highlighted cells reveals a much sparser network structure, with most of the upper-diagonal elements contributing little to the overall forecast variance of the entire network. The majority of small bilateral spillovers contributing less than 1\% of forecast variance are pruned. Only 40\% of the off-diagonal elements are selected by the information criterion. The remaining edges cluster heavily around the U.S.-Europe and Asia-Pacific blocks. The U.S. remains the largest number of outgoing connections, while markets such as Chile, Mexico, and Turkey become nearly isolated.

The sparse structure fundamentally changes the interpretation of the Diebold-Yilmaz network. The resulting network unveils a more economically meaningful topology of global linkages. Critical risk transmission channels are highlighted via the sparse network, for example, from the U.S. to European markets or from Hong Kong to Asian markets, while redundant or negligible pathways are suppressed. Much of the global integration is driven by a small number of dominant cross-market channels rather than universal contagion. This more parsimonious risk transmission structure makes systemic risk analysis more tractable: rather than monitoring hundreds of weak pairwise connections, regulators and investors can focus on the few dominant channels of global spillovers. The IN and OUT degrees reported at the bottom of Table \ref{tab:DY2009} provide additional insights into the directional structure of shock transmission. The OUT degree counts the non-zero elements in a given column and measures how many other markets are affected by shocks originating in that market. Conversely, the IN degree counts the number of non-zero entries in a row, representing the number of other markets that influence a given market. These results reveal the directional asymmetry of global stock market connectedness. The developed markets in the U.S. and Europe act as sources of return transmission, while emerging markets serve primarily as receivers of external shocks.

\subsection{S\&P 500 Sector Network}

The second example applies the sparsity-based connectedness framework to the S\&P 500 sector indices of the U.S. stock market. The 11 sector indices measure the performance of companies in the S\&P 500, with equal weights assigned to companies classified within the relevant sectors based on their GICS classification. We use daily returns from 8/11/2006 to 30/06/2025, totaling 4687 trading days. The Real Estate sector is excluded from the analysis because it was launched in 2016, leaving 10 sector indices in the network: Health Care, Financials, Communication Services, Energy, Materials, Industrials, Consumer Discretionary, Consumer Staples, Information Technology, and Utilities. We first remove the influence of overall market movements on sector indices by regressing each sector's daily return on the contemporaneous S\&P 500 market return, retaining the residuals for subsequent network analysis. This pre-filtering step ensures that the network analysis focuses on the sector-specific shock transmission rather than common shocks that simultaneously affect all sectors, such as macroeconomic news or global sentiment shifts. Similar factor-based approaches have been applied in the network literature, for example, \citet{AndoEtAl2022, AndoEtAl2024}.

\begin{sidewaystable}
\caption{S\&P 500 sectoral network \label{tab:SPsector}}
\centering
\footnotesize
\begin{tabular}{r C{1.6cm}C{1.6cm}C{1.6cm}C{1.6cm}C{1.6cm}C{1.6cm}C{1.6cm}C{1.6cm}C{1.6cm}C{1.6cm}}
\toprule \hline
& \multicolumn{10}{c}{From}\T\B \\ \cline{2-11}
& & & & \multicolumn{4}{c}{\textbf{8/11/2006\text{--}31/12/2015}}\T\B \\ \cline{5-8}
To  & Health. & Financ. & Comm. & Energy & Mater. & Indust. & Cs.Disc. & Cs.Stap. & I.T. & Util.\T\B \\ \hline 
Health.  & 83.27\sig & 4.80\sig & 0.57  & 1.30\sig & 0.94\sig & 0.14  & 0.17  & 6.98\sig & 0.96\sig & 0.89\sig\T \\ 
Financ.  & 4.45\sig & 78.19\sig & 0.13  & 1.97\sig & 0.97\sig & 0.62  & 9.65\sig & 1.92\sig & 0.67  & 1.44\sig \\ 
Comm.  & 0.63  & 0.77  & 91.93\sig & 0.08  & 0.19  & 0.20  & 0.82  & 1.74\sig & 0.60  & 3.05\sig \\ 
Energy  & 1.21\sig & 1.96\sig & 0.08  & 73.58\sig & 15.55\sig & 1.74\sig & 1.95\sig & 2.08\sig & 0.37  & 1.49\sig \\ 
Mater.  & 0.84\sig & 0.29  & 0.16  & 13.57\sig & 64.19\sig & 14.93\sig & 2.78\sig & 0.38  & 2.82\sig & 0.04  \\ 
Indust.  & 0.54  & 0.50  & 0.37  & 1.46\sig & 15.10\sig & 64.40\sig & 11.51\sig & 0.27  & 5.62\sig & 0.23  \\ 
Cs.Disc.  & 0.40  & 8.59\sig & 0.87\sig & 1.60\sig & 2.82\sig & 11.78\sig & 65.71\sig & 0.96\sig & 5.90\sig & 1.38\sig \\ 
Cs.Stap.  & 6.50\sig & 1.99\sig & 1.61\sig & 2.22\sig & 0.75  & 0.31  & 0.97\sig & 78.03\sig & 0.94\sig & 6.69\sig \\ 
I.T.  & 0.50  & 0.50  & 0.34  & 0.17  & 3.50\sig & 6.82\sig & 6.55\sig & 0.83\sig & 78.27\sig & 2.52\sig \\ 
Util.  & 0.81  & 1.67\sig & 2.81\sig & 1.65\sig & 0.04  & 0.28  & 1.70\sig & 6.60\sig & 2.73\sig & 81.70\sig\B \\ \hline 
& & & & \multicolumn{4}{c}{\textbf{4/01/2016\text{--}30/06/2025}}\T\B \\ \cline{5-8}
  & Health. & Financ. & Comm. & Energy & Mater. & Indust. & Cs.Disc. & Cs.Stap. & I.T. & Util.\T\B \\ \hline 
Health.  & 89.94\sig & 0.04  & 0.10  & 0.57  & 0.79  & 0.46  & 0.01  & 5.17\sig & 0.14  & 2.78\sig\T \\ 
Financ.  & 0.01  & 50.86\sig & 0.65\sig & 8.18\sig & 12.82\sig & 16.19\sig & 7.94\sig & 1.29\sig & 1.18\sig & 0.88\sig \\ 
Comm. & 0.06  & 1.12\sig & 88.04\sig & 0.70  & 1.29\sig & 1.20\sig & 5.67\sig & 1.49\sig & 0.13  & 0.29  \\ 
Energy  & 0.37  & 10.51\sig & 0.60  & 64.45\sig & 12.70\sig & 7.60\sig & 3.08\sig & 0.07  & 0.51  & 0.11  \\ 
Mater.  & 0.44  & 12.01\sig & 0.70\sig & 9.23\sig & 47.73\sig & 18.84\sig & 8.40\sig & 1.67\sig & 0.09  & 0.88\sig \\ 
Indust.  & 0.22  & 14.45\sig & 0.62\sig & 5.31\sig & 17.95\sig & 45.52\sig & 14.02\sig & 1.20\sig & 0.06  & 0.64\sig \\ 
Cs.Disc.  & 0.05  & 8.87\sig & 3.70\sig & 2.48\sig & 10.00\sig & 17.46\sig & 57.07\sig & 0.26  & 0.07  & 0.04  \\ 
Cs.Stap.  & 3.89\sig & 1.69\sig & 1.17\sig & 0.03  & 2.26\sig & 1.73\sig & 0.29  & 67.73\sig & 3.67\sig & 17.52\sig \\ 
I.T.  & 0.13  & 1.81\sig & 0.16  & 0.61  & 0.11  & 0.04  & 0.07  & 4.18\sig & 89.64\sig & 3.27\sig \\ 
Util.  & 2.13\sig & 1.26\sig & 0.24  & 0.03  & 1.31\sig & 0.93\sig & 0.05  & 18.74\sig & 3.05\sig & 72.26\sig\B \\ 
\hline \bottomrule 
\end{tabular}
\vspace{0.2cm}

\footnotesize
\parbox{21.75cm}{\textit{Notes}: The top panel reports the GFEVD network estimated using daily data from 8/11/2006 to 31/12/2015. The bottom panel reports the estimated network using daily data from 4/01/2016 to 30/06/2025. Cells shaded in \crule[purple!25]{0.6cm}{0.25cm} denote connectedness selected by the information criterion. The penalty term is determined using the POOF-MSFE procedure outlined in Section \ref{subsec:lambda}.}
\end{sidewaystable}

\begin{figure}[ht]
    \centering
    \caption{S\&P 500 sectoral network\label{fig:SPsector}} 
    \subfloat[8/11/2006\text{--}31/12/2015]
	{\includegraphics[width=0.5\textwidth]{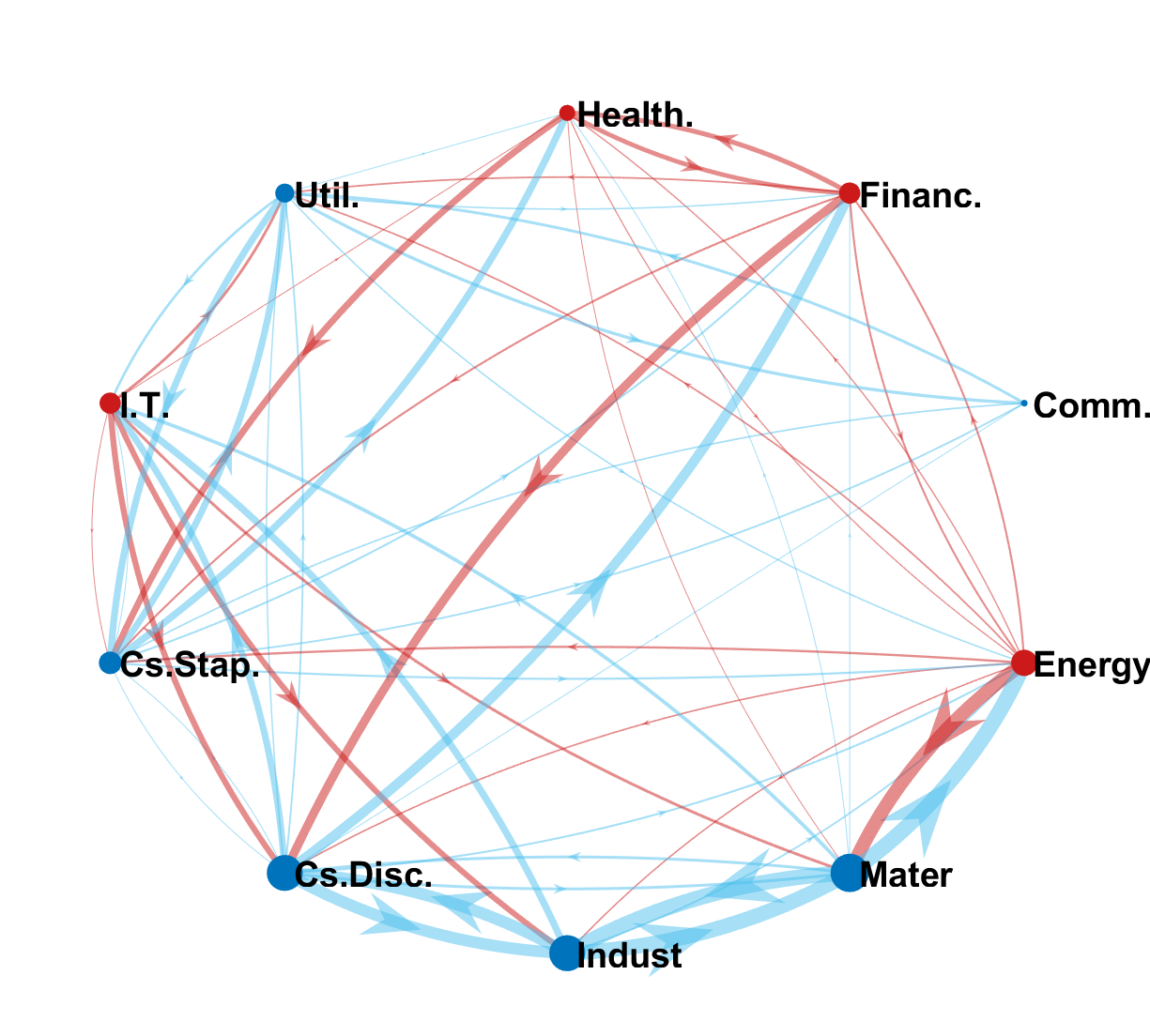}}
    \subfloat[4/01/2016\text{--}30/06/2025]
	{\includegraphics[width=0.5\textwidth]{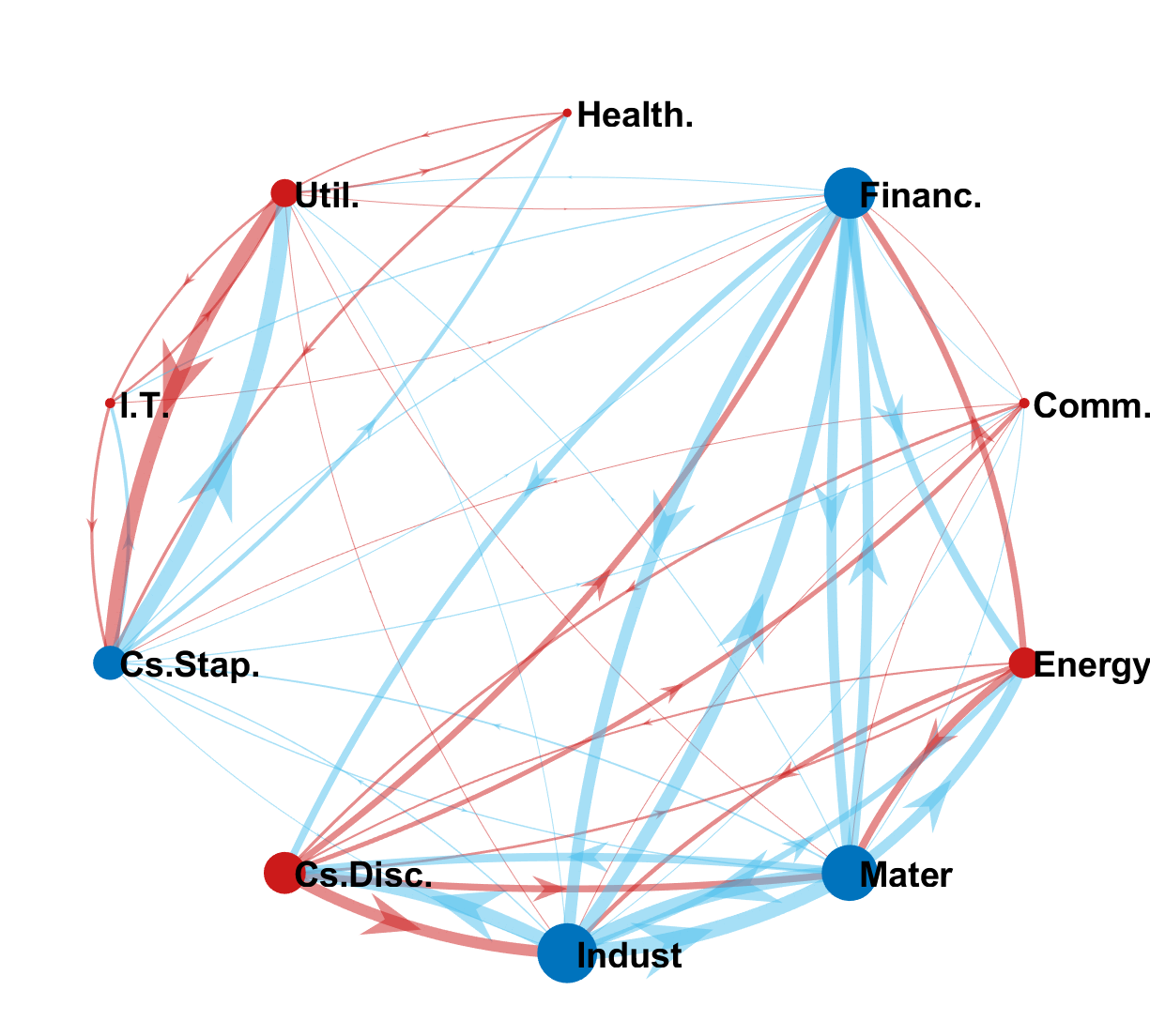}}
\footnotesize

\parbox{16.25cm}{\textit{Notes}: Panel (a) plots the network graph for the sample from 8/11/2006 to 31/12/2015. Panel (b) plots the network graph for the sample from 4/01/2016 to 30/06/2025. The size of the nodes is proportional to the total connectedness TIX+FIX of each node. Blue nodes represent sectors with net positive spillover (TIX$-$FIX$>$0) and red nodes represent sectors with net negative spillover (TIX$-$FIX$<$0). The color of the edges is the same as the color of the nodes from which the edges originate. The thickness of the edges is proportional to the strength of the pairwise connectedness.}
\end{figure}

We estimate a VAR(1) for the 10-variable sector network and use the GFEVD approach of \citet{DieboldYilmaz2014} with $H=10$ to construct the connectedness table. Table \ref{tab:SPsector} reports the estimated network structure. The entire sample is divided into two periods. The top panel of Table \ref{tab:SPsector} uses data from 8/11/2006 to the end of 2015, encompassing the Global Financial Crisis (GFC) and the subsequent recovery. The bottom panel uses data from the beginning of 2016 to 30/06/2025, covering the expansion prior to the COVID-19 pandemic and the booming stock market after the pandemic shock. Pairwise connections that are selected by the information criterion are highlighted in purple. The composition of dominant spillovers changes markedly between the two periods, indicating a structural evolution in the internal linkages of the U.S. equity market. Figure \ref{fig:SPsector} provides a visualization of the non-zero pairwise connections selected by the information criterion. Panel (a) plots the network graph for the sample from 8/11/2006 to 31/12/2015. Panel (b) plots the network graph for the sample from 4/01/2016 to 30/06/2025. The sum of TIX and FIX for each node is used as a measure of its connectedness with the rest of the network. The size of the node in Figure \ref{fig:SPsector} is proportional to this sum. Another metric of empirical interest is the difference between TIX and FIX, which is sometimes referred to as the net spillover index: NIX$=$TIX$-$FIX. Nodes with positive NIX are colored in blue in Figure \ref{fig:SPsector}, suggesting that these sectors transmit more shocks to others than they receive. Nodes with negative NIX are colored  in red, representing the net recipients of shocks. The color of the directional edges is the same as the color of the nodes from which the edges originate. The thickness of the edges is proportional to the strength of the directional connectedness.

In the first subsample, Health Care and Communication Services are the least connected sector, exhibiting a high own-variance share and few connections with other sectors. The Financials sector is a strong shock exporter, which is not surprising given the central role of financial institutions during the GFC. Energy also acts as a key transmitter, particularly with Materials and Industrials, consistent with the influence of commodity price fluctuations on cyclically sensitive sectors. In total, 56 out of the 90 pairwise connections are selected by the information criterion. The sparse network graph aligns with the macroeconomic narrative of the time, that the systemic risk concentrated in credit and commodity markets propagates through production-oriented sectors. In the second subsample, Health Care becomes more insulated from the rest of the U.S. equity market alongside the Information Technology sector. The Information Technology sector grows to be more self-driven; its disconnect from Materials, Industrials, and Consumer Discretionary coincides with the surge of large-cap technology firms in the S\&P 500 and the AI boom post-COVID. Financials, while still influential, also receive substantial spillovers from other sectors, making it a net spillover receiver. The bilateral feedback among these sectors stands in sharp contrast to the unidirectional dominance of Financials before 2016. The Financials sector appears to be strongly integrated with all other sectors except Health Care.

The comparison across the two periods reveals several important insights into the temporal evolution of the sparse network. First, because the sector returns are orthogonalized with respect to the aggregate S\&P 500 market return, the resulting network captures the idiosyncratic transmission, i.e., how shocks originated in one sector affect another after controlling for the overall market. The sparsity of the estimated network reflects the limited number of genuine direct connections among sectors after removing the common market component. Second, results in Table \ref{tab:SPsector} and Figure \ref{fig:SPsector} demonstrate that the dominant sources of transmission have shifted over time. From a systemic-risk perspective, monitoring should focus on the small subset of core sectors whose shocks propagate widely. Last but not least, the zero connections indicate that many sectors move largely independently, conditional on the market factor. This minimal connectedness can be used to diversify across sectors \citep[see, for example,][]{AntonakakisEtAl2019}.

\subsection{Commodity Volatility Connectedness Network}

To further demonstrate the flexibility of the sparsity-based connectedness framework, we examine volatility spillovers across major commodity futures. Commodities provide a distinctive setting for connectedness analysis, because they reflect both economic fundamentals and financial market forces. The network structure among commodities is expected to be considerably sparser than that of equities, as cross-market dependencies are typically concentrated within related commodity groups rather than spanning all categories. The analysis covers 24 commodity futures listed in Table \ref{tab:commodity}, grouped into five broad categories: Energy, Metals, Grains, Livestock, and Other Agriculture. The sample composition follows \citet{DieboldLiuYilmaz2017}, \citet{YangLiMiao2021} and \citet{DelleChiaieEtAl2022}.

\begin{table}[ht]
\centering
\caption{The list of commodity futures\label{tab:commodity}}
\footnotesize
\begin{tabular}{@{\hskip 0.75cm}l@{\hskip 1cm}l@{\hskip 1cm}l@{\hskip 1cm}l@{\hskip 0.75cm}}
\toprule \hline
\textbf{Category} & \textbf{Commodity} & \textbf{RIC} & \textbf{Exchange}\T\B \\ \hline
Energy  & (1) Light Sweet Crude Oil WTI & CLc1 & NYMEX\T \\
        & (2) Henry Hub Natural Gas     & NGc1 & NYMEX \\
        & (3) RBOB Gasoline             & RBc1 & NYMEX \\
        & (4) ULS Diesel (Heating Oil)  & HOc1 & NYMEX\B \\
Metals  & (1) Gold                      & GCc1 & CME\T \\
        & (2) Silver                    & SIc1 & CME \\
        & (3) High Grade Copper         & HGc1 & CME \\
        & (4) Nickel                  & CMNIc1 & LME \\
        & (5) Zinc                    & CMZNc1 & LME \\
        & (6) Aluminum                & CMALc1 & LME \\
        & (7) Lead                    & CMPBc1 & LME\B \\
Grains  & (1) Corn                      & Cc1  & CBOT\T \\
        & (2) Wheat                     & Wc1  & CBOT \\
        & (3) Soybean                   & Sc1  & CBOT \\
        & (4) Oats                      & Oc1  & CBOT\B \\
Livestock & (1) Live Cattle             & LCc1 & CME\T \\
          & (2) Feeder Cattle           & FCc1 & CME \\
          & (3) Lean Hogs               & LHc1 & CME\B \\
Other Agriculture & (1) Cocoa           & CCc1 & ICE\T \\
        & (2) Coffee                    & KCc1 & ICE \\
        & (3) Sugar No.11               & SBc1 & ICE \\
        & (4) Frozen Orange             & OJc1 & ICE \\
        & (5) Cotton NO.2               & CTc1 & ICE \\
        & (6) Lumber                   & LXRc1 & CME\B \\ \hline
\bottomrule
\end{tabular}
\end{table}

\begin{sidewaystable}
\caption{Commodity log-volatility network 2023\text{--}2025 \label{tab:RVcommodity}}
\centering
\scriptsize
\begin{tabular}{l rrrr| rrrrrrr| rrrr| rrr| rrrrrr}
\toprule \hline
& \multicolumn{24}{c}{From}\T\B \\ \cline{2-25}
    & \multicolumn{4}{c|}{Energy} & \multicolumn{7}{c|}{Metals} & \multicolumn{4}{c|}{Grains} & \multicolumn{3}{c|}{Livestock} & \multicolumn{6}{c}{Other Agriculture}\T\B \\
To  & (1) & (2) & (3) & (4) & (1) & (2) & (3) & (4) & (5) & (6) & (7) & (1) & (2) & (3) & (4) & (1) & (2) & (3) & (1) & (2) & (3) & (4) & (5) & (6)\T\B \\ \hline 
(1)  & 40.7\sig & 0.8\sig & 24.0\sig & 25.0\sig & 1.1\sig & 0.4  & 1.4\sig & 0.1  & 0.1  & 0.9\sig & 0.1  & 0.9\sig & 1.8\sig & 0.7  & 0.3  & 0.1  & 0.3  & 0.1  & 0.2  & 0.1  & 0.3  & 0.0  & 0.5  & 0.0 \T \\ 
(2)  & 2.4\sig & 65.0\sig & 1.5\sig & 2.0\sig & 0.5  & 0.2  & 0.6  & 0.3  & 0.2  & 6.0\sig & 0.7  & 4.5\sig & 0.5  & 1.2  & 1.1  & 2.0\sig & 1.9\sig & 3.8\sig & 0.9  & 0.4  & 0.1  & 0.8  & 2.2\sig & 1.2  \\ 
(3)  & 26.2\sig & 0.6  & 38.4\sig & 21.8\sig & 0.5  & 0.2  & 0.6  & 0.3  & 0.2  & 1.9\sig & 0.1  & 0.9\sig & 3.1\sig & 0.8\sig & 0.6  & 0.2  & 0.2  & 0.7  & 1.0\sig & 0.4  & 0.1  & 0.1  & 1.0\sig & 0.1  \\ 
(4)  & 27.0\sig & 1.4\sig & 21.6\sig & 41.4\sig & 0.3  & 0.1  & 0.4  & 0.1  & 0.2  & 0.8\sig & 0.1  & 0.9\sig & 1.6\sig & 0.5  & 0.2  & 0.2  & 0.2  & 0.4  & 1.5\sig & 0.2  & 0.2  & 0.1  & 0.3  & 0.1  \\ \hline
(1)  & 3.8\sig & 0.6  & 1.4\sig & 1.0  & 76.8\sig & 1.4\sig & 3.8\sig & 0.4  & 0.2  & 0.6  & 0.7  & 0.9  & 0.5  & 0.4  & 0.2  & 1.2  & 0.9  & 1.9\sig & 1.1  & 0.3  & 0.4  & 1.1  & 0.1  & 0.4  \\ 
(2)  & 0.6  & 0.7  & 0.6  & 0.5  & 1.9\sig & 83.2\sig & 4.3\sig & 0.2  & 0.5  & 0.5  & 0.4  & 0.8  & 0.2  & 0.6  & 0.1  & 0.4  & 0.4  & 0.5  & 1.0  & 0.8  & 0.2  & 0.3  & 0.4  & 0.9  \\ 
(3)  & 1.3  & 1.9\sig & 0.9  & 0.9  & 1.9\sig & 2.7\sig & 78.6\sig & 0.3  & 0.6  & 0.6  & 0.7  & 3.5\sig & 0.3  & 0.7  & 0.1  & 0.2  & 1.3  & 0.4  & 0.7  & 0.3  & 1.4  & 0.1  & 0.2  & 0.3  \\ 
(4)  & 0.7  & 0.7  & 1.4  & 1.6\sig & 1.0  & 0.3  & 0.5  & 76.3\sig & 2.1\sig & 3.3\sig & 2.0\sig & 0.6  & 1.2  & 0.6  & 0.3  & 0.1  & 0.3  & 1.4  & 0.4  & 2.4\sig & 0.3  & 0.3  & 1.7\sig & 0.7  \\ 
(5)  & 0.2  & 0.2  & 0.8  & 0.6  & 0.6  & 0.7  & 0.8  & 1.4\sig & 78.5\sig & 5.1\sig & 2.0\sig & 0.2  & 1.4\sig & 1.1  & 0.2  & 0.1  & 0.8  & 1.3  & 0.2  & 0.2  & 0.2  & 0.9  & 1.6\sig & 0.8  \\ 
(6)  & 1.1  & 1.8\sig & 2.2\sig & 0.6  & 0.8  & 1.1  & 0.4  & 1.3  & 4.9\sig & 69.9\sig & 2.0\sig & 0.2  & 2.6\sig & 1.6\sig & 0.1  & 0.8  & 0.6  & 2.1\sig & 0.1  & 0.2  & 0.3  & 0.4  & 3.9\sig & 1.1  \\ 
(7)  & 0.4  & 0.6  & 0.1  & 0.5  & 0.2  & 0.3  & 0.3  & 1.8\sig & 2.0\sig & 1.3  & 83.6\sig & 0.6  & 0.7  & 1.0  & 0.1  & 0.3  & 0.8  & 0.2  & 0.4  & 0.6  & 0.2  & 0.3  & 1.4  & 2.2\sig \\ \hline
(1)  & 1.8\sig & 2.5\sig & 1.6\sig & 1.6\sig & 0.7  & 0.5  & 1.8\sig & 0.1  & 0.1  & 0.4  & 0.5  & 63.8\sig & 6.0\sig & 8.8\sig & 2.0\sig & 1.4\sig & 0.8  & 0.3  & 2.4\sig & 0.2  & 0.6  & 0.3  & 1.1  & 0.5  \\ 
(2)  & 2.8\sig & 0.5  & 3.4\sig & 2.4\sig & 1.0  & 0.1  & 0.4  & 0.8  & 0.5  & 1.3\sig & 0.2  & 6.1\sig & 66.1\sig & 6.6\sig & 3.5\sig & 0.2  & 0.2  & 0.3  & 0.6  & 1.0  & 0.1  & 0.3  & 1.1  & 0.6  \\ 
(3)  & 1.5\sig & 1.2\sig & 1.4\sig & 1.2\sig & 0.7  & 0.2  & 0.7  & 0.2  & 0.5  & 0.6  & 0.3  & 10.0\sig & 8.5\sig & 66.2\sig & 2.8\sig & 0.0  & 0.1  & 0.2  & 0.1  & 0.2  & 0.6  & 0.2  & 0.9  & 1.6\sig \\ 
(4)  & 1.4\sig & 1.4\sig & 2.2\sig & 0.5  & 0.5  & 0.1  & 0.3  & 1.4\sig & 0.2  & 1.0  & 0.3  & 2.7\sig & 4.9\sig & 3.7\sig & 72.8\sig & 0.1  & 0.9  & 0.2  & 0.8  & 0.4  & 0.4  & 1.2  & 2.1\sig & 0.4  \\ \hline
(1)  & 0.1  & 2.0\sig & 0.8  & 0.3  & 2.1\sig & 0.2  & 0.3  & 0.2  & 0.2  & 2.4\sig & 0.4  & 0.8  & 0.2  & 0.0  & 0.1  & 76.5\sig & 9.9\sig & 0.3  & 0.8  & 0.1  & 0.4  & 1.4  & 0.5  & 0.1  \\ 
(2)  & 0.5  & 0.7  & 0.7  & 0.4  & 2.4\sig & 0.2  & 0.4  & 0.1  & 1.2  & 1.4\sig & 0.7  & 0.1  & 0.2  & 0.1  & 0.2  & 9.9\sig & 78.6\sig & 0.3  & 1.0  & 0.1  & 0.5  & 0.1  & 0.4  & 0.2  \\ 
(3)  & 0.2  & 0.3  & 0.5  & 0.1  & 2.8\sig & 0.1  & 2.3\sig & 0.6  & 1.7\sig & 3.9\sig & 0.5  & 0.2  & 0.9  & 0.8  & 0.0  & 0.9  & 0.9  & 79.5\sig & 1.1  & 0.1  & 0.9  & 0.2  & 0.4  & 1.1  \\ \hline
(1)  & 1.8\sig & 0.1  & 4.8\sig & 4.0\sig & 1.3  & 0.6  & 0.9  & 0.4  & 0.1  & 0.9  & 0.2  & 0.3  & 0.4  & 0.1  & 0.2  & 0.2  & 0.8  & 3.8\sig & 76.3\sig & 0.3  & 0.6  & 1.2  & 0.6  & 0.0  \\ 
(2)  & 0.4  & 0.4  & 0.6  & 0.2  & 0.1  & 0.1  & 0.2  & 1.6\sig & 0.3  & 0.2  & 0.4  & 1.0  & 1.2  & 0.1  & 0.4  & 0.2  & 0.3  & 0.6  & 2.5\sig & 88.5\sig & 0.3  & 0.1  & 0.2  & 0.1  \\ 
(3)  & 0.2  & 1.1  & 1.6\sig & 0.6  & 0.2  & 1.1  & 0.1  & 0.2  & 0.3  & 0.8  & 0.1  & 0.4  & 0.6  & 1.0  & 0.7  & 0.1  & 0.8  & 0.4  & 2.2\sig & 0.2  & 86.8\sig & 0.3  & 0.2  & 0.0  \\ 
(4)  & 0.4  & 0.4  & 0.1  & 0.1  & 1.4  & 0.5  & 0.4  & 0.9  & 0.1  & 1.1  & 1.1  & 0.3  & 0.2  & 0.4  & 1.3  & 0.9  & 0.3  & 1.4  & 1.5  & 0.9  & 0.7  & 84.7\sig & 0.1  & 0.7  \\ 
(5)  & 2.1\sig & 1.6\sig & 3.0\sig & 1.2  & 0.4  & 0.4  & 1.1  & 0.6  & 1.6\sig & 4.6\sig & 0.5  & 1.4  & 1.1  & 1.1  & 2.4\sig & 0.2  & 0.4  & 0.4  & 0.6  & 0.2  & 0.1  & 0.5  & 73.9\sig & 0.7  \\ 
(6)  & 0.3  & 0.6  & 0.2  & 0.5  & 0.7  & 0.4  & 0.1  & 0.8  & 0.6  & 0.5  & 0.2  & 1.1  & 0.4  & 0.6  & 0.1  & 0.2  & 0.8  & 0.3  & 0.6  & 0.0  & 0.1  & 0.9  & 0.1  & 90.1\sig\B \\ \hline
IN   & 8  & 9 & 8  & 7 & 5 & 2 & 4 & 6 & 5 & 8  & 3 & 10 & 7 & 8 & 8 & 4 & 3 & 4 & 4 & 2 & 2 & 0 & 6 & 0\T \\
OUT  & 10 & 9 & 12 & 8 & 6 & 2 & 5 & 4 & 5 & 11 & 3 & 8  & 8 & 5 & 4 & 3 & 2 & 4 & 5 & 1 & 0 & 0 & 6 & 2\B \\
\hline \bottomrule 
\end{tabular}
\vspace{0.2cm}

\footnotesize
\parbox{24cm}{\textit{Notes}: Sample period 3 January 2023 \text{--} 30 September 2025.}
\end{sidewaystable}

For each commodity, we construct daily realized volatility (RV) from 5-minute intraday prices. The sample for the estimation reported in Table \ref{tab:RVcommodity} covers the period from 3 January 2023 to 30 September 2025. We set $p=2$ for the VAR model based on the autocorrelation structure of the log-RV and compute the GFEVD with $H=10$ following \citet{DieboldYilmaz2014}. The information criterion developed in Section \ref{sec:model} is applied to retain only those connections that contribute significantly to forecast variance. The resulting sparse GFEVD matrix is reported in Table \ref{tab:RVcommodity}. The non-zero pairwise connections chosen by the information criterion are highlighted in purple. Table \ref{tab:RVcommodity} shows that the resulting volatility connectedness network among commodities is quite sparse. Of the 552 possible off-diagonal connections, only 123 pairwise spillovers are retained by the information criterion, representing 22\% of all potential links. This low density confirms that volatility transmission across commodity markets is limited and highly selective once redundant relationships are eliminated. 

The sparse GFEVD network in Table \ref{tab:RVcommodity} exhibits a clear block-diagonal structure, suggesting strong volatility interactions within rather than across each of the commodity categories. The four energy contracts form the most tightly connected group. Bidirectional spillovers are strong, particularly between crude oil, gasoline, and heating oil, reflecting the refining chain that links these markets. Energy volatilities also transmit weak but notable shocks to the grains group, consistent with energy's role in transportation costs and biofuel production. Metals show moderate within-group connectivity. There is a division between the precious metals, which are more commonly used as hedging instruments in the financial market, and the base metals traded on the London Metal Exchange (LME), which are primarily driven by industrial demand. The four grain contracts show economically meaningful within-group volatility linkages. These connections mirror their shared exposure to input cost and weather shocks. Livestock and other agricultural commodities have the weakest overall connectivity. Only a handful of links survive the selection, suggesting that their volatility remains largely idiosyncratic.

The bottom rows of Table \ref{tab:RVcommodity} summarize the IN and OUT degrees of each commodity. Energy commodities stand out as the primary net transmitters of volatility, accounting for 39 of the 123 retained spillovers. Their central role aligns with their function as a universal input to production and transport. Yet, the low network density suggests that energy volatility shocks propagate selectively across commodity categories. Metals and grains play the secondary roles, acting both as recipients and limited transmitters. In particular, aluminum and wheat exhibit higher out- than in-degrees, suggesting that industrial and agricultural commodities can occasionally amplify volatility transmission across sectors. Livestock and other agricultural categories are often net recipients of volatility, except for cocoa. Their low out-degree confirms that shocks originating in these markets rarely influence other markets.

Overall, the commodity volatility network exemplifies a naturally sparse and modular system. Only a small number of economically meaningful connections drive cross-market volatility transmission, while the vast majority of potential spillovers are effectively zero. This example demonstrates the effectiveness of the information-criterion approach in filtering out spurious comovements, yielding a concise, economically interpretable network representation.

\section{Conclusion}\label{sec:conclusion}

This paper proposes a novel, data-driven framework for uncovering the sparse structure of the Diebold-Yilmaz financial networks. By reformulating the FEVD through a regression perspective, we bridge the gap between network analysis and model selection. We develop information criteria based on FEVD and GFEVD to systematically distinguish economically meaningful spillover channels from statistical noise, addressing the limitations of dense connectedness measures that often obscure the true network topology. We also propose a data-driven procedure to select the penalty parameter in the information criteria using pseudo out-of-sample forecast performance.

Our extensive Monte Carlo simulations demonstrate that the proposed methods perform well in finite samples. More importantly, they remain robust to approximately sparse networks and heavy-tailed error distributions. The data-driven tuning procedure is shown to effectively balance model fit and parsimony, ensuring consistent recovery of the active set of linkages. Empirically, our applications to global stock markets, S\&P 500 sectoral indices, and commodity futures challenge the conventional \citetalias{DieboldYilmaz2014} methods, which consistently yield dense financial interconnectedness. Instead, our empirical results demonstrate that many financial networks often exhibit significant sparsity.

Although this paper does not take a position on the broader debate regarding the use of FEVD with orthogonal shocks versus GFEVD for network construction, our simulation evidence sheds light on the relative performance of the two approaches in different settings. For practitioners implementing the proposed framework, the choice between FEVD and GFEVD should be guided by the specific analytical objective. The GFEVD-based criterion is better suited to settings where comprehensive recovery of the network topology is desired or when variables lack a clear structural ordering, as it consistently delivers higher detection rates for active connections and preserves weaker but potentially relevant spillovers. In contrast, the FEVD-based criterion is preferable when the goal is to isolate dominant transmission channels in networks with clustered structures. By enforcing sparsity more aggressively and exhibiting greater robustness to heavy-tailed errors, it provides a parsimonious representation that limits false positives in noisy environments.

The proposed framework provides a transparent and computationally tractable approach to identifying sparse network structures. By explicitly addressing the trade-off between model fit and sparsity, it moves the network analysis beyond mechanically dense connectedness measures and toward representations that more clearly isolate economically meaningful transmission channels. This feature is particularly valuable for risk monitoring and policy analysis, where overly dense networks can obscure the sources of systemic vulnerability. We leave it for future research.

 


\newpage
\begin{singlespace}
    \bibliographystyle{agsm}
    \bibliography{references}
\end{singlespace}

\newpage
\appendix
\setcounter{page}{1}

\begin{center}
{\LARGE\textbf{Appendix}} 
\par\end{center}

\section{Regarding Assumption \ref{A:A1} and Proofs}\label{sec:appd1}

Appendix \ref{APP:low_level_condition} shows the convergence
rate of $\hat{\varphi}$ from using either Cholesky decomposition
or eigenvalue decomposition. Appendix \ref{APP:Cholesky} presents
the procedure of the Cholesky Decomposition to facilitate the proof
of the convergence rate of $\hat{\varphi}.$ Appendix \ref{APP:proof}
proves Theorem \ref{TH:main}.

\subsection{Regarding $\hat{\varphi}$ }

\label{APP:low_level_condition}

Lemmas \ref{LE:sufficient} and \ref{LE:sufficient-1} show the results
when using Cholesky decomposition and eigenvalue decomposition, respectively.

\begin{lemma}\label{LE:sufficient}Suppose Assumptions \ref{A:model}
and \ref{A:m} hold. $\hat{\varphi}$ is obtained via Cholesky decomposition.
Then $\hat{\varphi}_{ij}^{2}$, $i,j=1,2,...,m$, satisfy 
\[
\hat{\varphi}_{ij}^{2}-\varphi_{ij}^{2}=O_{P}\left(T^{-1/2}\right).
\]

\end{lemma}

\noindent \textbf{Proof of Lemma \ref{LE:sufficient}.} Let $\Phi_{l,ij}$
denote the $(i,j)$-th element of $\mathbf{\Phi}_{l}$, and $\hat{\Phi}_{l,ij}$
is similarly defined. By Assumptions \ref{A:model} and \ref{A:m}
and the property of ordinary least squares estimators, 
\[
\hat{\Phi}_{l,ij}-\Phi_{l,ij}=O_{P}\left(T^{-1/2}\right),l=1,2,...,p,i,j=1,2,...,m.
\]

Note that 
\begin{equation}
\hat{a}\hat{b}+\hat{c}\hat{d}-ab-cd=O_{P}\left(T^{-1/2}\right),\label{eq:abcd}
\end{equation}
if $\hat{a},\hat{b},\hat{c},\textrm{ and }\hat{d}$ converge to $a,b,c,\textrm{ and }d$
at the speed of $T^{-1/2},$ respectively. Since each element of $\hat{\mathbf{\Psi}}_{l}$
is some simple summations and multiplications of elements in $\mathbf{\hat{\Phi}}_{1},...,\mathbf{\hat{\Phi}}_{l}$,
the above implies that the $(i,j)$-th element of $\hat{\mathbf{\Psi}}_{l}$
also converges at the rate of $T^{-1/2}.$ That is 
\begin{equation}
\hat{\Psi}_{l,ij}-\Psi_{l,ij}=O_{P}\left(T^{-1/2}\right),l=1,2,...,\textrm{ and }i,j=1,2,...,m.\label{eq:Psilij}
\end{equation}

Each element of $\hat{\mathbf{\Sigma}}$ converges to the corresponding
true at the rate of $T^{-1/2}$ for the same reason and thanks to
the finite fourth moment of $\boldsymbol{\varepsilon}_{t}$ in Assumption
\ref{A:model}.

We similarly let $P_{ij}$ and $\hat{P}_{ij}$ denote the $(i,j)$-th
element of $\mathbf{P}$ and $\mathbf{\hat{P}},$ respectively. For
the convergence of $\hat{P}_{ij}$, we note that 
\begin{equation}
\frac{\hat{a}\hat{b}}{\hat{c}}-\frac{ab}{c}=O_{P}\left(T^{-1/2}\right)\textrm{ and }\sqrt{\hat{c}}-\sqrt{c}=O_{P}\left(T^{-1/2}\right),\label{eq:abchat}
\end{equation}
if $\hat{a},\hat{b},\textrm{ and }\hat{c}$ converge to $a,b,\textrm{ and }c$
at the speed of $T^{-1/2},$ respectively, and $c$ is bounded away
from zero. From the Cholesky decomposition in Appendix \ref{APP:Cholesky},
$P_{ij}$ is computed from a series of operations in (\ref{eq:abchat}).
Further, $c$ only appears in diagonals of $\mathbf{P}$. By the positive
definiteness of $\mathbf{\Sigma}$ in Assumption \ref{A:model}, the
diagonals of $\mathbf{\Sigma}$ must be strictly positive. Consequently,
the diagonals of $\mathbf{P}$ must be bounded away from zero . As
a result, the condition to apply (\ref{eq:abchat}) for elements in
$\mathbf{P}$ and $\mathbf{\hat{P}}$ are satisfied, and thus 
\begin{equation}
\hat{P}_{ij}-P_{ij}=O_{P}\left(T^{-1/2}\right),i,j=1,2,...,m.\label{eq:Pij}
\end{equation}

Finally, note that $\hat{\varphi}_{ij}^{2}$ is a series of operations
as in (\ref{eq:abcd}) on elements in $\mathbf{\hat{\Psi}}_{l}\textrm{ and }\hat{\mathbf{P}}.$
Therefore, (\ref{eq:Psilij}) and (\ref{eq:Pij}) imply that 
\[
\hat{\varphi}_{ij}^{2}-\varphi_{ij}^{2}=O_{P}\left(T^{-1/2}\right),i,j=1,2,...,m.
\]
\hfill{}$\blacksquare$

\begin{lemma}\label{LE:sufficient-1}Suppose Assumptions \ref{A:model}
and \ref{A:m} hold. $\hat{\varphi}$ is obtained via eigenvalue decomposition.
Then $\hat{\varphi}_{ij}^{2}$, $i,j=1,2,...,m$, satisfy 
\[
\hat{\varphi}_{ij}^{2}-\varphi_{ij}^{2}=O_{P}\left(T^{-1/2}\right).
\]

\end{lemma}

\noindent\textbf{Proof of Lemma \ref{LE:sufficient-1}.} Take $\mathbf{\Sigma}$
in (\ref{eq:sigmaPP'1}) to illustrate, eigenvalues of $\mathbf{\Sigma}$
are $P_{jj}^{2}$. Although the procedures of Cholesky and eigenvalue
decomposition are different, the eigenvalues obtained from both procedures
are identical. We show in the proof of Lemma \ref{LE:sufficient}
that $\hat{P}_{jj}-P_{jj}=O_{P}\left(T^{-1/2}\right)$, thus $\hat{P}_{jj}^{2}-P_{jj}^{2}=O_{P}\left(T^{-1/2}\right)$.
In other words, eigenvalues from both procedures converge to the true
at the rate of $T^{-1/2}.$

Given the above, it is sufficient to show eigenvectors converge to the
true at the rate of $T^{-1/2}$ to complete the proof. Suppose for
eigenvalue $\lambda_{j}$, the corresponding eigenvector is $\boldsymbol{\xi}_{j}$
with $\left\Vert \boldsymbol{\xi}_{j}\right\Vert =1.$ Then 
\begin{equation}
\left(\mathbf{\Sigma}-\lambda_{j}\mathbf{I}_{m}\right)\boldsymbol{\xi}_{j}=0.\label{eq:linearEquation}
\end{equation}
Suppose other eigenvalues differ from $\lambda_{j}$ (other cases
can be similarly proved). We can equivalently normalize one of the
nonzero elements in $\boldsymbol{\xi}_{j}$ to $1$ if it is positive
and $-1$ if it is negative, say $\tilde{\boldsymbol{\xi}}_{j}$.
With that normalized element removed, denote the vector as $\tilde{\boldsymbol{\xi}}_{j,-1}$.
We can move the corresponding column of $\mathbf{\Sigma}$ to the
right hand side of (\ref{eq:linearEquation}), say $\boldsymbol{b}$,
and we obtain $\mathbf{\tilde{\Sigma}}$. Since $\mathbf{\Sigma}-\lambda_{j}\mathbf{I}$
is singular, we can remove one redundant row of $\mathbf{\tilde{\Sigma}}$
to make it full rank, because other eigenvalues differ from $\lambda_{j}$.
Say, we have 
\[
\left(\mathbf{\mathbf{\tilde{\Sigma}}}_{-1}-\lambda_{j}\mathbf{I}_{m-1}\right)\tilde{\xi}_{j,-1}=\mathbf{b}{}_{-1},
\]
where $\tilde{\Sigma}_{-1}-\lambda_{j}I$ is full rank. We can solve
$\tilde{\xi}_{j,-1}$ from the linear equations as 
\[
\tilde{\xi}_{j,-1}=\left(\tilde{\Sigma}_{-1}-\lambda_{j}\mathbf{I}_{m-1}\right)^{-1}\mathbf{b}{}_{-1}=\frac{\textrm{\texttt{adj}}\left(\tilde{\Sigma}_{-1}-\lambda_{j}\mathbf{I}_{m-1}\right)\cdot\mathbf{b}{}_{-1}}{\textrm{\texttt{det}}\left(\mathbf{\tilde{\Sigma}}_{-1}-\lambda_{j}\mathbf{I}_{m-1}\right)},
\]
where \texttt{adj} is the adjugate, and \texttt{det} is the determinant.

Note that \texttt{adj} and \texttt{det} only involve calculations
in (\ref{eq:abcd}). Therefore, the sample counterparts from eigenvalue
decomposition of $\textrm{\texttt{adj}}\left(\mathbf{\tilde{\Sigma}}_{-1}-\lambda_{j}\mathbf{I}_{m-1}\right)\cdot\mathbf{b}_{-1}$
and $\textrm{\texttt{det}}\left(\tilde{\Sigma}_{-1}-\lambda_{j}\mathbf{I}_{m-1}\right)$
converge to the true at the rate of $T^{-1/2}.$ Since $\textrm{\texttt{det}}\left(\mathbf{\tilde{\Sigma}}_{-1}-\lambda_{j}\mathbf{I}_{m-1}\right)$
is nonzero, we can apply the result in (\ref{eq:abchat}) such that
the sample counterpart of $\tilde{\xi}_{j,-1}$ converge to the true
at the rate of $T^{-1/2},$ as desired. \hfill{}$\blacksquare$

\subsection{The Cholesky Decomposition}\label{APP:Cholesky}

To illustrate, we explain the Cholesky decomposition of a $3\times3$
positive definite matrix $\mathbf{\Sigma}=\left\{ \sigma_{ij}\right\} _{3\times3}$
in the following. If the equation 

\begin{align}
\mathbf{\Sigma} & =\mathbf{P}\mathbf{P}'=\begin{pmatrix}P_{11} & 0 & 0\\
P_{21} & P_{22} & 0\\
P_{31} & P_{32} & P_{33}
\end{pmatrix}\begin{pmatrix}P_{11} & P_{21} & P_{31}\\
0 & P_{22} & P_{32}\\
0 & 0 & P_{33}
\end{pmatrix}\label{eq:sigmaPP'}\\
 & =\begin{pmatrix}P_{11}^{2} & P_{21}P_{11} & P_{31}P_{11}\\
P_{21}P_{11} & P_{21}^{2}+P_{22}^{2} & P_{31}P_{21}+P_{32}P_{22}\\
P_{31}P_{11} & P_{31}P_{21}+P_{32}P_{22} & P_{31}^{2}+P_{32}^{2}+P_{33}^{2}
\end{pmatrix}\nonumber 
\end{align}
is written out, the following is obtained: 
\[
\mathbf{P}=\begin{pmatrix}\sqrt{\sigma_{11}} & 0 & 0\\
\sigma_{21}/P_{11} & \sqrt{\sigma_{22}-P_{21}^{2}} & 0\\
\sigma_{31}/P_{11} & (\sigma_{32}-P_{31}P_{21})/P_{22} & \sqrt{\sigma_{33}-P_{31}^{2}-P_{32}^{2}}
\end{pmatrix}.
\]
For a general $m\times m$ positive definite matrix 
\begin{equation}
\mathbf{\Sigma}=\left\{ \sigma_{ij}\right\} _{m\times m}=\mathbf{P}\mathbf{P}',\label{eq:sigmaPP'1}
\end{equation}
we have
\[
P_{jj}=\sqrt{\sigma_{jj}-\sum_{k=1}^{j-1}P_{jk}^{2}},j=1,2,...,m,\textrm{ and }
\]
\[
P_{ij}=\frac{1}{P_{jj}}\left(\sigma_{ij}-\sum_{k=1}^{j-1}P_{ik}P_{jk}\right),\text{ for }j<i\leq m,j=1,2,...,m.
\]

\subsection{Proof of Theorem \ref{TH:main}}

\label{APP:proof}

\noindent \textbf{Proof.} Without loss of generality, we show $\Pr\left(\hat{k}=k^{*}\right)\rightarrow1$
in two steps, 
\[
\Pr\left(\textrm{IC}_{\textrm{FEVD}}^{H}\left(k^{*},\lambda_{T}\right)<\textrm{IC}_{\textrm{FEVD}}^{H}\left(k^{*}-1,\lambda_{T}\right)\right)\rightarrow1
\]
and 
\[
\Pr\left(\textrm{IC}_{\textrm{FEVD}}^{H}\left(k^{*},\lambda_{T}\right)<\textrm{IC}_{\textrm{FEVD}}^{H}\left(k^{*}+1,\lambda_{T}\right)\right)\rightarrow1.
\]
Other cases can be similarly proved. Denote $C_{\min}\equiv\min_{(i,j)\in\mathscr{M}}\varphi_{ij}^{2}$.
By the definition of $\mathscr{M}$ and the fact that $k^{*}$ (from
Assumption \ref{A:m}) is a fixed number, 
\[
C_{\min}=\min_{(i,j)\in\mathscr{M}}\varphi_{ij}^{2}>0.
\]
We first claim the following: 
\begin{equation}
\Pr\left(\min_{(i,j)\in\mathscr{M}}\hat{\varphi}_{ij}^{2}\leq\frac{C_{\min}}{2}\leq\max{}_{(i,j)\in\mathscr{M}}\hat{\varphi}_{ij}^{2}\right)\rightarrow0,\label{eq:claim}
\end{equation}
whose proof is deferred to the end.

The second part of the result, $\Pr\left(\mathscr{\hat{M}}=\mathscr{M}\right)\rightarrow1$,
is a direct result of $\Pr\left(\hat{k}=k^{*}\right)\rightarrow1$,
(\ref{eq:claim}) and $\left|\mathscr{M}\right|=k^{*}$.

\noindent\textbf{Step 1: }Denote the event 
\[
\mathscr{\hat{E}}=\left\{ \min_{(i,j)\in\mathscr{M}}\hat{\varphi}_{ij}^{2}>\frac{C_{\min}}{2}>\max{}_{(i,j)\in\mathscr{M}}\hat{\varphi}_{ij}^{2}\right\} .
\]
The claim implies $\Pr\left(\mathscr{\hat{E}}\right)\rightarrow1.$
We first show the result by assuming that $\mathscr{\hat{E}}$ holds.
Conditional on $\mathscr{\hat{E}},$ 
\begin{equation}
\hat{\varphi}^{\left(k^{*}\right)}>C_{\min}/2,\label{eq:phi(k)}
\end{equation}
because $\left|\mathscr{M}\right|=k^{*}$. Then, 
\begin{align*}
 & \textrm{IC}_{\textrm{FEVD}}^{H}\left(k^{*}-1,\lambda_{T}\right)-\textrm{IC}_{\textrm{FEVD}}^{H}\left(k^{*},\lambda_{T}\right)\\
= & 2T\left[\log\left(m-\sum_{l=1}^{k^{*}-1}\left(\hat{\varphi}^{\left(l\right)}\right)^{2}\right)-\log\left(m-\sum_{l=1}^{k^{*}}\left(\hat{\varphi}^{\left(l\right)}\right)^{2}\right)\right]-\lambda_{T}\\
= & 2T\log\left(1+\frac{\left(\hat{\varphi}^{\left(k^{*}\right)}\right)^{2}}{m-\sum_{l=1}^{k^{*}}\left(\hat{\varphi}^{\left(l\right)}\right)^{2}}\right)-\lambda_{T}\\
\geq & 2T\log\left(1+\frac{C_{\min}}{2m}\right)-\lambda_{T}>0,\textrm{ after some large }T,
\end{align*}
where the last line holds by (\ref{eq:phi(k)}) and $\lambda_{T}/T\rightarrow0.$
Therefore, 
\[
\Pr\left(\textrm{IC}_{\textrm{FEVD}}^{H}\left(k^{*},\lambda_{T}\right)<\textrm{IC}_{\textrm{FEVD}}^{H}\left(k^{*}-1,\lambda_{T}\right)\right)\rightarrow1
\]
because $\mathscr{\hat{E}}$ implies the above event, and $\Pr\left(\mathscr{\hat{E}}\right)\rightarrow1.$

\noindent\textbf{Step 2: }We also show the result conditional on
$\mathscr{\hat{E}}.$ Note 
\begin{align*}
 & \textrm{IC}_{\textrm{FEVD}}^{H}\left(k^{*}+1,\lambda_{T}\right)-\textrm{IC}_{\textrm{FEVD}}^{H}\left(k^{*},\lambda_{T}\right)\\
= & 2T\left[\log\left(m-\sum_{l=1}^{k^{*}+1}\left(\hat{\varphi}^{\left(l\right)}\right)^{2}\right)-\log\left(m-\sum_{l=1}^{k^{*}}\left(\hat{\varphi}^{\left(l\right)}\right)^{2}\right)\right]+\lambda_{T}\\
= & 2T\log\left(1-\frac{\left(\hat{\varphi}^{\left(k^{*}+1\right)}\right)^{2}}{m-\sum_{l=1}^{k^{*}}\left(\hat{\varphi}^{\left(l\right)}\right)^{2}}\right)+\lambda_{T}\text{.}
\end{align*}
Conditional on $\mathscr{\hat{E}},$ $\hat{\varphi}^{\left(k^{*}+1\right)}$
corresponds to an element in $\mathscr{M}^{c}$. Recall that $\varphi_{ij}^{2}=0$
for $\left(i,j\right)\in\mathscr{M}^{c}$. Using the same logic for the claim at the end of the proof, we can show that 
\begin{equation}
\max_{\left(i,j\right)\in\mathscr{M}^{c}}\left|\hat{\varphi}_{ij}\right|=O_{P}\left(T^{-1/2}\right).\label{eq:maxphij}
\end{equation}
In addition, by the positive definiteness of $\mathbf{\Sigma}$ (in
other words, $\textrm{Var}\left(\varepsilon_{ti}\right)>0$ for $i=1,2,...,m$), there exists a positive $C,$ 
\begin{equation}
m-\sum_{l=1}^{k^{*}}\left(\hat{\varphi}^{\left(l\right)}\right)^{2}>C>0\label{eq:IC_denominator}
\end{equation}
with very high probability. (\ref{eq:maxphij}) and (\ref{eq:IC_denominator})
imply that 
\begin{equation}
2T\log\left(1-\frac{\left(\hat{\varphi}^{\left(k^{*}+1\right)}\right)^{2}}{m-\sum_{l=1}^{k^{*}}\left(\hat{\varphi}^{\left(l\right)}\right)^{2}}\right)\approx\frac{2T\left(\hat{\varphi}^{\left(k^{*}+1\right)}\right)^{2}}{m-\sum_{l=1}^{k^{*}}\left(\hat{\varphi}^{\left(l\right)}\right)^{2}}=2TO_{P}\left(\left(\hat{\varphi}^{\left(k^{*}+1\right)}\right)^{2}\right)=O_{P}\left(1\right).\label{eq:step2approx}
\end{equation}
By $\lambda_{T}\rightarrow\infty,$ we must have 
\[
\Pr\left(\left.\textrm{IC}_{\textrm{FEVD}}^{H}\left(k^{*}+1,\lambda_{T}\right)-\textrm{IC}_{\textrm{FEVD}}^{H}\left(k^{*},\lambda_{T}\right)>0\right|\mathscr{\hat{E}}\right)\rightarrow1.
\]
The above holds unconditionally thanks to $\Pr\left(\mathscr{\hat{E}}\right)\rightarrow1,$
using 
\[
\Pr\left(A\right)\geq\Pr\left(A|\mathscr{\hat{E}}\right)\Pr\left(\mathscr{\hat{E}}\right)\rightarrow1,
\]
if $\Pr\left(A|\mathscr{\hat{E}}\right)\rightarrow1.$

Steps 1 and 2 complete the proof. We now show the claim in (\ref{eq:claim}).

\noindent\textbf{Proof of the Claim:} $\Pr\left(\min_{(i,j)\in\mathscr{M}}\hat{\varphi}_{ij}^{2}\leq\max{}_{(i,j)\in\mathscr{M}^c}\hat{\varphi}_{ij}^{2}\right)\rightarrow0.$

By definition, $\mathscr{M\cup M}^{c}=\left\{ \left.\left(i,j\right)\right|:i,j=1,2,...,m\right\} $.
Thus, $\left|\mathscr{M\cup M}^{c}\right|=m^{2}$, which is finite
by Assumption \ref{A:m}. From Assumption \ref{A:A1}, the finiteness
of $\left|\mathscr{M\cup M}^{c}\right|$ implies the uniform convergence
of $\hat{\varphi}_{ij}^{2}$ for $\left(i,j\right)\in\left|\mathscr{M\cup M}^{c}\right|$.\footnote{Take $\hat{b}_{j},j=1,2,$ as an example. Suppose $\hat{b}_{j}-b_{j}=O_{P}\left(T^{-1/2}\right).$
For any small positive $\epsilon_{1}$ and $\epsilon_{2}$, then there
exist a $N_{j},$such that for $T>N_{j},$ 
\[
\Pr\left(\left|\hat{b}_{j}-b_{j}\right|\geq\epsilon_{2}\right)<\epsilon_{1},j=1,2.
\]
Denote $N=\max\left\{ N_{1},N_{2}\right\} .$ Then for $T>N,$ 
\[
\Pr\left(\max_{j=1,2}\left|\hat{b}_{j}-b_{j}\right|\geq\epsilon_{2}\right)\leq\sum_{j=1}^{2}\Pr\left(\left|\hat{b}_{j}-b_{j}\right|\geq\epsilon_{2}\right)<2\epsilon_{1}.
\]
Since $\epsilon_{1}$ is any small positive constant, we can simply
let $\epsilon=2\epsilon_{1},$ and we show the uniform convergence
of $\hat{b}_{j},j=1,2.$ The above hold for any finite number of $\hat{b}_{j},$
e.g., $j=1,2,....,m^{2}.$ However, the above does not apply to a
diverging number of $\hat{b}_{j}.$} That is, for any small $\epsilon$, there exist a $N,$ such that
for all $T>N$, 
\[
\Pr\left(\max_{i,j\in\mathscr{M\cup M}^{c}}\left|\hat{\varphi}_{ij}^{2}-\varphi_{ij}^{2}\right|\geq\frac{C_{\min}}{2}\right)<\epsilon.
\]
The above implies that 
\[
\Pr\left(\min_{i,j\in\mathscr{M}}\hat{\varphi}_{ij}^{2}\leq\frac{C_{\min}}{2}\right)<\epsilon,
\]
and 
\[
\Pr\left(\max_{i,j\in\mathscr{M}^{c}}\hat{\varphi}_{ij}^{2}\geq\frac{C_{\min}}{2}\right)<\epsilon,
\]
by the definition of $\mathscr{M\textrm{ and }M}^{c}$. The above two inequalities imply that 
\[
\Pr\left(\min_{(i,j)\in\mathscr{M}}\hat{\varphi}_{ij}^{2}\leq\frac{C_{\min}}{2}\leq\max{}_{(i,j)\in\mathscr{M}^c}\hat{\varphi}_{ij}^{2}\right)<\epsilon,
\]
as desired. \hfill{}$\blacksquare$
\newpage
\section{Additional Simulation Results}\label{sec:tabs_and_figs}
\setcounter{figure}{0}
\setcounter{table}{0}
\renewcommand{\thefigure}{B.\arabic{figure}}
\renewcommand{\thetable}{B.\arabic{table}}

\begin{figure}[p]
\centering
\caption{The distribution of the selected constant $c^\ast$ for DGPs L1 and L2 with FEVD\label{fig:fevd_tuning}}
\subfloat[DGP L1]{\includegraphics[width=1\textwidth]{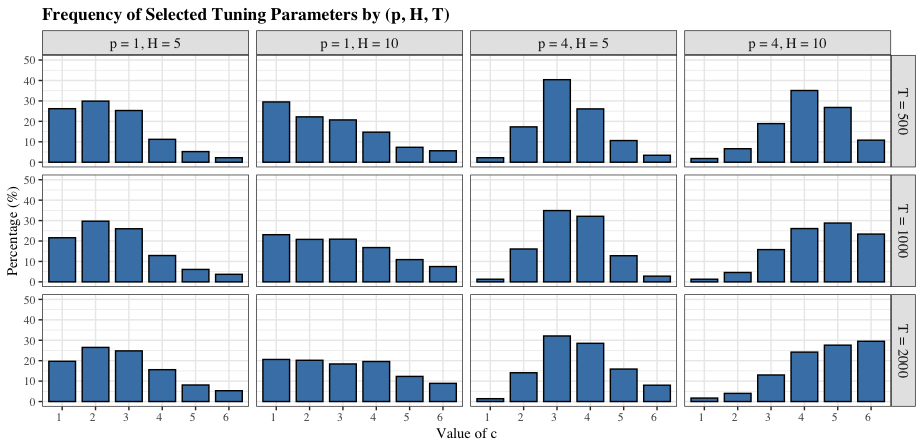}} \\ \vspace{0.5cm}
\subfloat[DGP L2]{\includegraphics[width=1\textwidth]{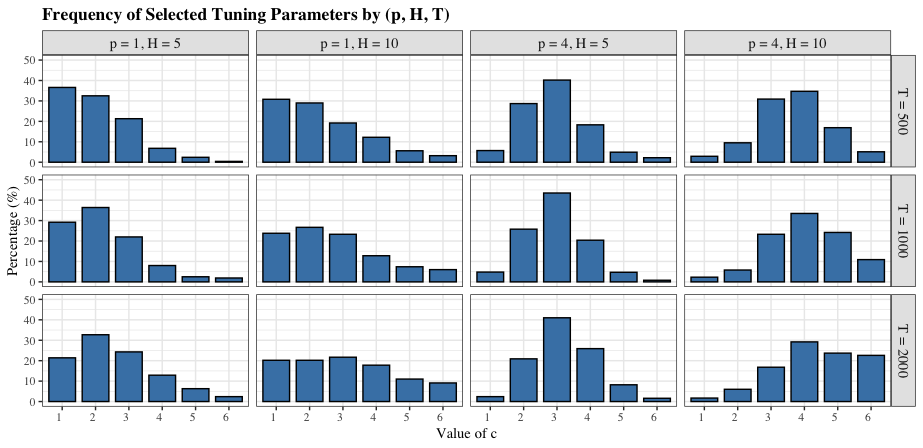}}
\end{figure}

Table \ref{tab:sim_large_FEVD} reports the correct discovery rates for large networks with 20 nodes using Cholesky decomposition in the FEVD. Figure \ref{fig:fevd_tuning} presents the distribution of the selected constant $c^\ast$ in the tuning parameter $\lambda_T$ for the two large networks L1 and L2 when FEVD is used.

\begin{table}[p]
  \centering
  \begin{threeparttable}
  \caption{Correct discovery rates for large networks with FEVD\label{tab:sim_large_FEVD}}
  \setlength{\tabcolsep}{4pt} 
  \begin{tabular*}{\textwidth}{@{\extracolsep{\fill}} lc *{3}{S[table-format=1.3]} *{3}{S[table-format=1.3]} *{3}{S[table-format=1.3]}}
    \toprule \hline
    & & \multicolumn{3}{c}{$H = 1$} & \multicolumn{3}{c}{$H = 5$} & \multicolumn{3}{c}{$H = 10$}\T \\ \cmidrule(lr){3-5} \cmidrule(lr){6-8} \cmidrule(lr){9-11}
    DGP & $T$ & {CDR$_1$} & {CDR$_0$} & {CDR$_a$} & {CDR$_1$} & {CDR$_0$} & {CDR$_a$} & {CDR$_1$} & {CDR$_0$} & {CDR$_a$} \\ \midrule
    
    \multicolumn{2}{c}{$p = 1$} \\ \cmidrule(lr){1-2}
    \multirow{3}{*}{L1} 
      & 500   & 0.484 & 0.930 & 0.888 & 0.836 & 0.953 & 0.931 & 0.892 & 0.930 & 0.923 \\
      & 1000  & 0.543 & 0.938 & 0.900 & 0.905 & 0.967 & 0.955 & 0.938 & 0.953 & 0.950 \\
      & 2000  & 0.597 & 0.944 & 0.911 & 0.946 & 0.976 & 0.970 & 0.965 & 0.964 & 0.964 \\ \addlinespace
    \multirow{3}{*}{L2} 
      & 500   & 0.421 & 0.927 & 0.859 & 0.814 & 0.941 & 0.907 & 0.877 & 0.922 & 0.910 \\
      & 1000  & 0.476 & 0.938 & 0.876 & 0.892 & 0.960 & 0.942 & 0.932 & 0.949 & 0.945 \\
      & 2000  & 0.529 & 0.947 & 0.891 & 0.938 & 0.975 & 0.965 & 0.963 & 0.965 & 0.964 \\ \addlinespace
    \multirow{3}{*}{L3} 
      & 500   & 0.723 & 0.930 & 0.927 & 0.864 & 0.980 & 0.976 & 0.882 & 0.961 & 0.959 \\
      & 1000  & 0.774 & 0.938 & 0.936 & 0.912 & 0.981 & 0.979 & 0.921 & 0.967 & 0.966 \\
      & 2000  & 0.810 & 0.945 & 0.943 & 0.939 & 0.982 & 0.981 & 0.948 & 0.969 & 0.969 \\ \addlinespace
    \multirow{3}{*}{L4} 
      & 500   & \text{--} & 0.927 & 0.927 & \text{--} & 0.972 & 0.972 & \text{--} & 0.937 & 0.937 \\
      & 1000  & \text{--} & 0.937 & 0.937 & \text{--} & 0.975 & 0.975 & \text{--} & 0.937 & 0.937 \\
      & 2000  & \text{--} & 0.943 & 0.943 & \text{--} & 0.974 & 0.974 & \text{--} & 0.941 & 0.941 \\ \midrule
    
    \multicolumn{2}{c}{$p = 4$} \\ \cmidrule(lr){1-2}
    \multirow{3}{*}{L1} 
      & 500   & 0.511 & 0.928 & 0.889 & 0.647 & 0.964 & 0.904 & 0.793 & 0.949 & 0.919 \\
      & 1000  & 0.576 & 0.936 & 0.902 & 0.814 & 0.975 & 0.945 & 0.905 & 0.969 & 0.956 \\
      & 2000  & 0.632 & 0.944 & 0.915 & 0.909 & 0.982 & 0.968 & 0.960 & 0.976 & 0.973 \\ \addlinespace
    \multirow{3}{*}{L2} 
      & 500   & 0.451 & 0.928 & 0.864 & 0.588 & 0.940 & 0.846 & 0.758 & 0.925 & 0.880 \\
      & 1000  & 0.513 & 0.935 & 0.879 & 0.779 & 0.956 & 0.909 & 0.887 & 0.956 & 0.938 \\
      & 2000  & 0.571 & 0.943 & 0.893 & 0.886 & 0.975 & 0.951 & 0.951 & 0.972 & 0.966 \\ \addlinespace
    \multirow{3}{*}{L3} 
      & 500   & 0.725 & 0.925 & 0.922 & 0.921 & 0.993 & 0.992 & 0.950 & 0.987 & 0.986 \\
      & 1000  & 0.789 & 0.934 & 0.933 & 0.974 & 0.995 & 0.995 & 0.984 & 0.986 & 0.986 \\
      & 2000  & 0.826 & 0.944 & 0.943 & 0.991 & 0.995 & 0.995 & 0.995 & 0.986 & 0.986 \\ \addlinespace
    \multirow{3}{*}{L4} 
      & 500   & \text{--} & 0.915 & 0.915 & \text{--} & 0.993 & 0.993 & \text{--} & 0.982 & 0.982 \\
      & 1000  & \text{--} & 0.932 & 0.932 & \text{--} & 0.995 & 0.995 & \text{--} & 0.984 & 0.984 \\
      & 2000  & \text{--} & 0.936 & 0.936 & \text{--} & 0.995 & 0.995 & \text{--} & 0.983 & 0.983 \\
    \hline \bottomrule
  \end{tabular*}%
  \begin{tablenotes}[flushleft]
    \footnotesize
    \item[]\textit{Notes}: This table reports the CDRs for 20-dimensional VAR($p$) models. L1 features a network with an active set size of $\left| \mathscr{M} \right|=72$, comprising one group of 8 nodes, one group of 4, two groups of 2, and 4 isolated nodes. L2 features a network with $\left| \mathscr{M} \right|=102$, comprising one group of 10 nodes, one group of 4, and 6 isolated nodes. L3 features a network with an active set size of $\left| \mathscr{M} \right|=10$, comprising 5 groups of 2 nodes and 10 isolated nodes. L4 features a null network with $\left| \mathscr{M} \right|=0$, representing a fully disconnected structure where all nodes are isolated.
  \end{tablenotes}
  \end{threeparttable}
\end{table}%

\begin{table}[p]
  \centering
  \begin{threeparttable}
  \caption{Selection measures for 20-node approximately sparse networks with GFEVD}\label{tab:sim_dense_gfevd}
  \setlength{\tabcolsep}{4pt} 
  \begin{tabular*}{\textwidth}{@{\extracolsep{\fill}} lc *{3}{S[table-format=1.3]} *{3}{S[table-format=1.3]} *{3}{S[table-format=1.3]}}
    \toprule \hline
    & & \multicolumn{3}{c}{$H = 1$} & \multicolumn{3}{c}{$H = 5$} & \multicolumn{3}{c}{$H = 10$}\T \\ \cmidrule(lr){3-5} \cmidrule(lr){6-8} \cmidrule(lr){9-11}
    DGP & $T$ & SP   & VL$_a$ & VL$_{o}$ & SP   & VL$_a$ & VL$_{o}$ & SP   & VL$_a$ & VL$_{o}$ \\ \midrule
    \multicolumn{2}{c}{$p = 1$} \\ \cmidrule(lr){1-2}
    \multirow{3}{*}{D1} 
          & 500   & 0.411 & 0.015 & 0.061 & 0.524 & 0.037 & 0.065 & 0.470 & 0.031 & 0.050 \\
          & 1000  & 0.325 & 0.008 & 0.034 & 0.373 & 0.020 & 0.035 & 0.325 & 0.017 & 0.027 \\
          & 2000  & 0.248 & 0.004 & 0.017 & 0.237 & 0.010 & 0.018 & 0.196 & 0.008 & 0.013 \\ \addlinespace
    \multirow{3}{*}{D2} 
          & 500   & 0.404 & 0.015 & 0.052 & 0.474 & 0.033 & 0.051 & 0.425 & 0.025 & 0.037 \\
          & 1000  & 0.315 & 0.008 & 0.028 & 0.342 & 0.018 & 0.028 & 0.297 & 0.014 & 0.020 \\
          & 2000  & 0.244 & 0.004 & 0.014 & 0.217 & 0.009 & 0.014 & 0.177 & 0.007 & 0.010 \\ \midrule
    \multicolumn{2}{c}{$p = 4$} \\  \cmidrule(lr){1-2}
    \multirow{3}{*}{D1} 
          & 500   & 0.395 & 0.014 & 0.055 & 0.562 & 0.046 & 0.101 & 0.410 & 0.032 & 0.060 \\
          & 1000  & 0.303 & 0.007 & 0.028 & 0.397 & 0.023 & 0.049 & 0.266 & 0.015 & 0.029 \\
          & 2000  & 0.227 & 0.003 & 0.012 & 0.248 & 0.011 & 0.023 & 0.149 & 0.007 & 0.013 \\ \addlinespace
    \multirow{3}{*}{D2} 
          & 500   & 0.380 & 0.013 & 0.042 & 0.518 & 0.044 & 0.083 & 0.382 & 0.028 & 0.047 \\
          & 1000  & 0.288 & 0.006 & 0.021 & 0.375 & 0.021 & 0.040 & 0.245 & 0.013 & 0.022 \\
          & 2000  & 0.219 & 0.003 & 0.009 & 0.233 & 0.009 & 0.017 & 0.141 & 0.006 & 0.009 \\
    \hline \bottomrule
  \end{tabular*}
  \begin{tablenotes}[flushleft]
    \footnotesize
    \item[]\textit{Notes}: This table presents the selection measures SP, VL$_a$ and VL$_o$ for 20-dimensional VAR($p$) models. DGPs D1 and D2 are adapted from L1 and L2, respectively, by replacing all zero elements in the coefficient and covariance matrices with small non-zero random numbers.
  \end{tablenotes}
  \end{threeparttable}
\end{table}

Table \ref{tab:sim_dense_gfevd} reports the selection proportion and the variance loss measures for 20-node dense networks using GFEVD.

\begin{figure}[p]
\centering
\caption{The distribution of the selected constant $c^\ast$ for DGPs H1 and H2 with FEVD\label{fig:fevd_tuning_H}}
\subfloat[DGP H1]{\includegraphics[width=1\textwidth]{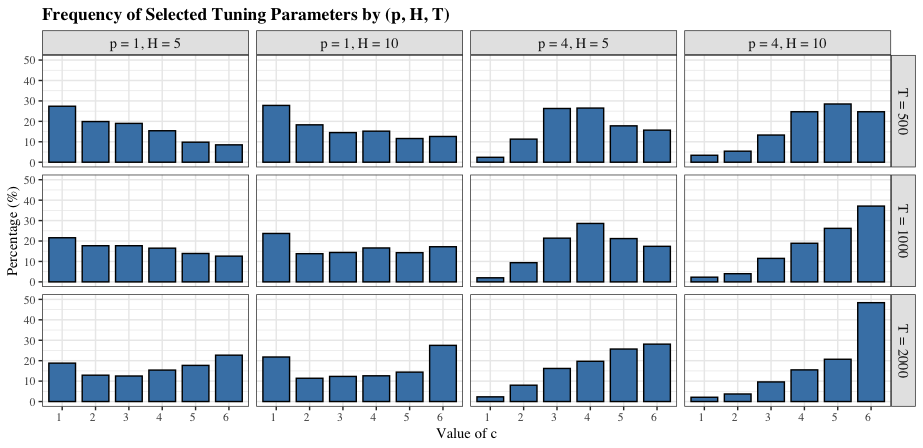}} \\ \vspace{0.5cm}
\subfloat[DGP H2]{\includegraphics[width=1\textwidth]{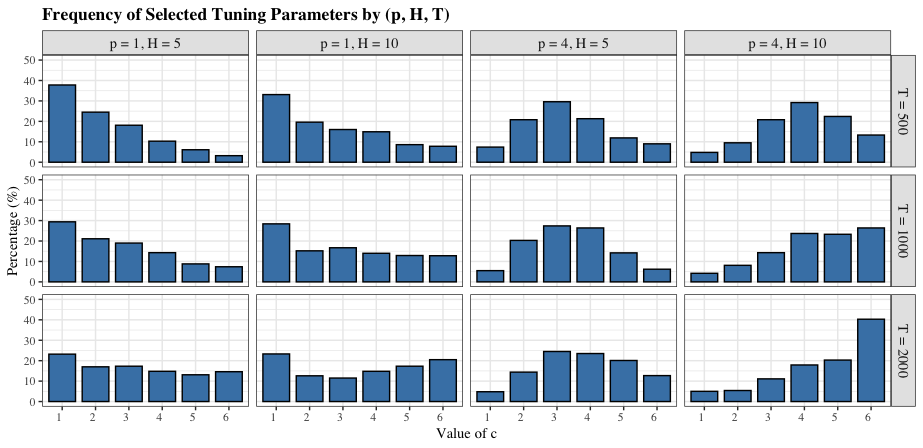}}
\end{figure}

\begin{figure}[p]
\centering
\caption{The distribution of the selected constant $c^\ast$ for DGPs H1 and H2 with GFEVD\label{fig:gfevd_tuning_H}}
\subfloat[DGP H1]{\includegraphics[width=1\textwidth]{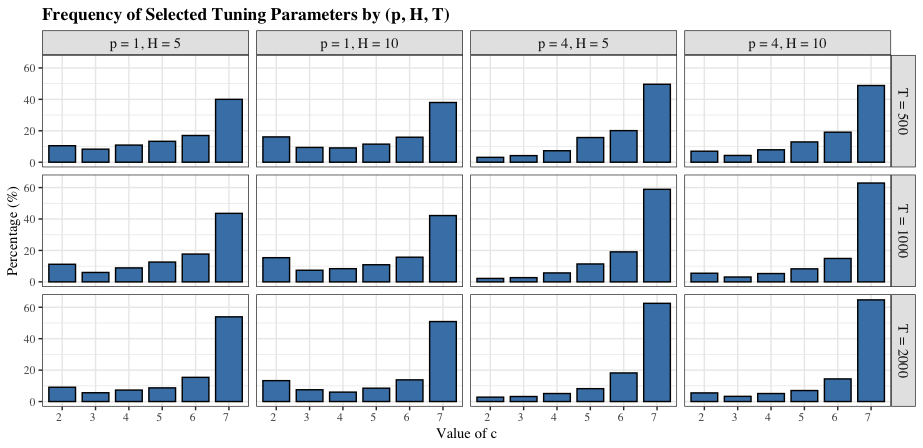}} \\ \vspace{0.5cm}
\subfloat[DGP H2]{\includegraphics[width=1\textwidth]{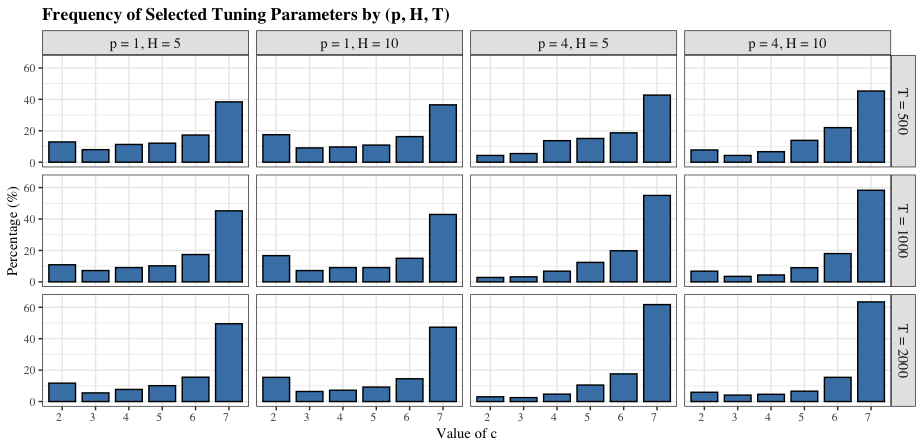}}
\end{figure}

Table \ref{tab:sim_heavy_tail_fevd} tabulates the correct discovery rates for 20-node networks with non-Gaussian errors when FEVD is used. Figures \ref{fig:fevd_tuning_H} and \ref{fig:gfevd_tuning_H} depict the distributions of the selected constant $c^\ast$ in the tuning parameter $\lambda_T$ for 20-node networks with non-Gaussian errors. Figure \ref{fig:fevd_tuning_H} uses FEVD, and Figure \ref{fig:gfevd_tuning_H} uses GFEVD.

\begin{table}[p]
  \centering
  \begin{threeparttable}
  \caption{Correct discovery rates for 20-node networks with non-Gaussian errors and FEVD}\label{tab:sim_heavy_tail_fevd}
  \setlength{\tabcolsep}{4pt} 
  \begin{tabular*}{\textwidth}{@{\extracolsep{\fill}} lc *{3}{S[table-format=1.3]} *{3}{S[table-format=1.3]} *{3}{S[table-format=1.3]}}
    \toprule \hline
    & & \multicolumn{3}{c}{$H = 1$} & \multicolumn{3}{c}{$H = 5$} & \multicolumn{3}{c}{$H = 10$}\T \\ \cmidrule(lr){3-5} \cmidrule(lr){6-8} \cmidrule(lr){9-11}
    DGP & $T$ & {CDR$_1$} & {CDR$_0$} & {CDR$_a$} & {CDR$_1$} & {CDR$_0$} & {CDR$_a$} & {CDR$_1$} & {CDR$_0$} & {CDR$_a$} \\ \midrule
    \multicolumn{2}{c}{$p = 1$} \\ \cmidrule(lr){1-2}
    \multirow{3}{*}{H1} 
        & 500   & 0.521 & 0.844 & 0.814 & 0.804 & 0.893 & 0.876 & 0.868 & 0.866 & 0.866 \\
        & 1000  & 0.581 & 0.839 & 0.815 & 0.876 & 0.910 & 0.904 & 0.922 & 0.886 & 0.893 \\
        & 2000  & 0.633 & 0.839 & 0.819 & 0.924 & 0.924 & 0.924 & 0.955 & 0.898 & 0.909 \\ \addlinespace
    \multirow{3}{*}{H2} 
        & 500   & 0.465 & 0.849 & 0.797 & 0.792 & 0.869 & 0.848 & 0.859 & 0.847 & 0.850 \\
        & 1000  & 0.518 & 0.849 & 0.805 & 0.868 & 0.894 & 0.887 & 0.922 & 0.871 & 0.885 \\
        & 2000  & 0.575 & 0.845 & 0.809 & 0.920 & 0.913 & 0.915 & 0.955 & 0.893 & 0.909 \\ \midrule
    
    \multicolumn{2}{c}{$p = 4$} \\ \cmidrule(lr){1-2}
    \multirow{3}{*}{H1} 
        & 500   & 0.556 & 0.843 & 0.816 & 0.587 & 0.941 & 0.874 & 0.758 & 0.911 & 0.882 \\
        & 1000  & 0.623 & 0.835 & 0.815 & 0.771 & 0.946 & 0.913 & 0.891 & 0.931 & 0.923 \\
        & 2000  & 0.674 & 0.832 & 0.817 & 0.879 & 0.954 & 0.940 & 0.951 & 0.940 & 0.942 \\ \addlinespace
    \multirow{3}{*}{H2} 
        & 500   & 0.496 & 0.852 & 0.804 & 0.543 & 0.907 & 0.810 & 0.726 & 0.879 & 0.838 \\
        & 1000  & 0.566 & 0.844 & 0.806 & 0.745 & 0.917 & 0.871 & 0.872 & 0.906 & 0.897 \\
        & 2000  & 0.618 & 0.844 & 0.814 & 0.859 & 0.933 & 0.913 & 0.943 & 0.919 & 0.925 \\
    \hline \bottomrule
  \end{tabular*}%
  \begin{tablenotes}[flushleft]
    \footnotesize
    \item[]\textit{Notes}: This table presents the CDRs for 20-dimensional VAR($p$) models with heavy-tailed errors. DGPs H1 and H2 are adapted from L1 and L2, respectively, by replacing Gaussian errors with Student-$t$ errors.
  \end{tablenotes}
  \end{threeparttable}
\end{table}%

\end{document}